\documentclass[11pt]{article}

\usepackage{jheppub}
\usepackage{amsmath,amssymb}
\usepackage{tikz-cd}
\usepackage{multicol}
\usepackage{mathrsfs}
\usepackage{circuitikz}

\newcommand{\be}{\begin{equation}}
\newcommand{\ee}{\end{equation}}
\newcommand{\bea}{\begin{eqnarray}}
\newcommand{\eea}{\end{eqnarray}}

\newcommand{\ba}{\begin{aligned}}
\newcommand{\ea}{\end{aligned}}
\newcommand{\p}{\partial}

\numberwithin{equation}{section}

\title{Light non-vacuum BMS$_3$ blocks at order $1/c_M$}

\author[a,b]{Boyang Yu}

\affiliation[a]{School of Mathematics and Maxwell Institute for Mathematical Sciences, University of Edinburgh, Edinburgh EH9 3FD, UK}
\affiliation[b]{Center for High Energy Physics, Peking University, No.5 Yiheyuan Rd, Beijing 100871, P.R. China}

\emailAdd{v1byu33@ed.ac.uk}
\abstract{
We derive a closed expression for the $1/c_M$ correction to highest-weight
BMS$_3$ blocks with four light external primaries of generic weights and
non-vacuum exchange. With $c_L$ and all operator weights held fixed, we
prove that the global sector and descendants containing exactly one
non-global generator $L_{-n}$ or $M_{-n}$ with $n\geq2$ suffice through this
order. This truncation reduces the intrinsic calculation to global blocks
and their derivatives, whose contributions we resum for nonzero exchanged
boost weight $\xi_p$. For pairwise identical external operators, we
independently reproduce the correction through the introduction of boundary gravitons in
the normalized three-point functions in the shadow representation, while
keeping the exchanged two-point function fixed. The analogous prescription
also reproduces the known non-vacuum Virasoro correction at order $1/c$.
For the same external configuration, a formal ultra-relativistic 
contraction of Virasoro blocks with one flipped chiral representation
provides a further check: it yields the same correction, and we prove
that contraction commutes with extraction of the $1/c_M$ coefficient at
each fixed descendant level. The degenerate exchange with $\xi_p=0$
requires a separate quotient construction and is not obtained as a
smooth limit of the generic block, which we discuss in the appendix.
}

\begin{document}
\maketitle

\section{Introduction}
\label{sec:introduction}

Conformal blocks separate the constraints of symmetry from the dynamical
data of a conformal field theory.  Each block collects the contribution of
an exchanged primary and its descendants, providing the basic ingredients
of the conformal bootstrap
\cite{Belavin:1984vu,Rattazzi:2008pe,Poland:2018epd}.
In two dimensions, Virasoro symmetry determines these contributions in
terms of the central charge and operator dimensions, but summing the
infinitely many descendants remains difficult.  Recursion relations
\cite{Zamolodchikov:1984eqp,Zamolodchikov:1987avt,Perlmutter:2015iya} and combinatorial
expansions associated with the AGT correspondence
\cite{Alday:2009aq,Alba:2010qc} give systematic access to the
blocks, while large central charge provides a regime in which explicit
analytic results can be obtained.

The large-$c$ expansion of Virasoro blocks is particularly well developed.
With all external and exchanged dimensions fixed, the leading term is the
global conformal block
\cite{Dolan:2003hv,Perlmutter:2015iya}. The first
correction for light non-vacuum exchange has been computed by Wilson line and descendant methods
\cite{Fitzpatrick:2016mtp,Hikida:2018dxe,Bombini:2018jrg}.
When two external dimensions instead scale with $c$, the heavy-light block
describes propagation in the background created by the heavy operators
\cite{Fitzpatrick:2014vua,Fitzpatrick:2015zha}.
Corrections to this limit and higher orders in the vacuum block have also
been computed
\cite{Beccaria:2015shq,Fitzpatrick:2015dlt,Chen:2016cms}.
When all dimensions scale with $c$, the semiclassical exponentiation of the block can instead be established using the oscillator
representation \cite{Besken:2019jyw}.  These distinct limits illustrate how the
organization of descendants controls both the leading answer and its
corrections.

For holographic theories, this expansion also probes quantum gravity in
AdS$_3$~\cite{Brown:1986nw,Maldacena:1997re}.  Global and semiclassical
blocks admit descriptions using geodesic Witten diagrams and worldlines
\cite{Hijano:2015zsa,Hijano:2015rla,Hijano:2015qja},
while Chern--Simons Wilson lines provide a route to their gravitational
corrections
\cite{Besken:2016ooo,Fitzpatrick:2016mtp,Besken:2017fsj,
Hikida:2017ehf,Hikida:2018dxe}.
A complementary approach describes the vacuum Virasoro block through
fluctuating boundary reparametrizations
\cite{Haehl:2018izb,Cotler:2018zff,Haehl:2019eae,Nguyen:2021jja}.
For non-vacuum exchange, the shadow formalism expresses the global block
through three-point functions and the exchanged two-point pairing
\cite{Ferrara:1972uq,Simmons-Duffin:2012juh}.  A proposed
extension of the reparametrization description to non-vacuum blocks combines reparametrized
three-point functions through a shadow transform \cite{Nguyen:2022xsw}. This suggests a concrete test: whether
boundary reparametrizations of the three-point functions reproduce the first correction beyond
the global block.

An analogous problem arises in three-dimensional flat holography, where
the asymptotic symmetry algebra is the centrally extended BMS$_3$ algebra
\cite{Barnich:2006av,Barnich:2010eb}.
Its relation to Galilean conformal symmetry
\cite{Bagchi:2009my,Bagchi:2009pe,Bagchi:2010zz}
and conformal Carroll symmetry~\cite{Duval:2014uva}
motivates developing conformal methods without Lorentz invariance.
The choice of representation is essential: highest-weight modules and
induced representations~\cite{Barnich:2014kra} are different, and an
algebraic contraction alone does not identify their blocks.
Here we study highest-weight BMS$_3$ blocks, defined by inserting the
projector onto an exchanged highest-weight module into a four-point
function.

The BMS bootstrap established this construction and its crossing equations
\cite{Bagchi:2016geg,Bagchi:2017cpu}.
Subsequent work developed global blocks, harmonic analysis and shadow
transforms~\cite{Chen:2020vvn,Chen:2022cpx}, as well as flat
space worldline constructions~\cite{Hijano:2017eii}.
Semiclassical blocks have been obtained by monodromy and oscillator methods
\cite{Hijano:2018nhq,Ammon:2020wem}, and more recently through
coordinate transformations adapted to heavy states~\cite{Hao:2025naz}.
On the gravity side, the boundary action provides a description of identity
blocks in terms of boundary gravitons \cite{Merbis:2019wgk}.
These results motivate determining the first correction beyond the global
block for non-vacuum exchange, and testing whether the same coefficient
can be obtained from a dressed shadow integral.

In this work, we compute this correction for four light highest-weight
primaries with independent weights $(\Delta_i,\xi_i)$, exchanging a
non-vacuum primary of weights $(\Delta_p,\xi_p)$.
The primaries considered here are singlets, with no boost multiplet at the top of their
modules.  We take $c_M\to\infty$ with the other central charge $c_L$ and all
weights fixed.
For $\xi_p\neq0$, the reduced block has the expansion
\begin{equation}
 \mathcal F_p(x,t)=\mathcal F_p^{(0)}(x,t)
 +\frac{1}{c_M}\mathcal F_p^{(1)}(x,t)+O(c_M^{-2}),
 \label{eq:intro-expansion}
\end{equation}
where $(x,t)$ are the BMS cross ratios and $\mathcal F_p^{(0)}$ is the
global block.  Our main result is a closed expression for
$\mathcal F_p^{(1)}$, including the sum over all descendant levels, in the
normalization fixed by the primary contribution to the OPE.  At this
order the coefficient is independent of $c_L$.

The central technical step is a truncation of the descendant calculation.
We assign a degree to each Poincar\'e--Birkhoff--Witt basis state by counting
its lowering generators $L_{-n}$ and $M_{-n}$ with $n\geq2$.
We prove that the degree-$s$ diagonal Gram block has a nondegenerate
leading term of order $c_M^s$ and that this scaling survives block
orthogonalization.  Consequently, only degrees zero and one contribute
through order $c_M^{-1}$.
We resolve degree one into pairs of global quasi-primary states and their
global descendants.  The action of $M_0$ on each pair is a rank-two Jordan
block, which produces shifted global blocks and derivatives with respect
to $\xi_p$ in the answer.  Their coefficients are fixed by three-point
functions, and the remaining sum can be performed in closed form.

For identical operators within each external pair, we give an independent prescription using boundary gravitons.  We transform the two normalized
three-point functions in the global shadow integral and contract their
linear responses using the boundary graviton correlator of
\cite{Merbis:2019wgk}, while holding the exchanged two-point pairing fixed.
The calculation reproduces the linearized correction from the intrinsic calculation.
  The analogous Virasoro calculation reproduces the known
light non-vacuum correction, providing a concrete test of the idea proposed
by Nguyen.  Both agreements concern only the first subleading order and a nonlinear prescription for the full block is left for further investigation.

A further check follows from an ultra-relativistic (UR) contraction of the
known Virasoro correction for the same external specialization.
We combine two chiral projectors, with one representation flipped so that
the contraction reaches a highest-weight BMS module
\cite{Bagchi:2019unf,Hao:2021urq}.
The closed expressions agree, and we prove that contraction commutes with
extraction of the $c_M^{-1}$ coefficient at every fixed descendant level.

The non-vacuum   quotient at $\xi_p=0$ requires a separate
calculation \cite{Chen:2022jhx}.
After removing the null submodule, we determine its global block and a closed expression for the first correction, assuming
$2\Delta_p\notin\mathbb Z_{\leq0}$.  These results are not obtained by
substituting $\xi_p=0$ into the nonzero $\xi_p$ answer.
The  external or exchanged BMS multiplets are outside
our scope.  The resulting formulas supply subleading data for the BMS
bootstrap and a benchmark for gravitational descriptions of non-vacuum
exchange.  Extending the dressing calculation beyond linear order and
adapting the descendant expansion to heavy states remain open problems.

Section \ref{sec:setup} reviews the BMS$_3$ algebra and the construction of BMS blocks.
Sections \ref{sec:degree} and \ref{sec:intrinsic} prove the truncation and
compute the correction intrinsically.  Sections \ref{sec:dressing} and \ref{sec:ur}
give the gravitational dressing and ultra-relativistic comparisons.
The appendices contain the three-point coefficients, the quasi-primary
construction, the $\xi_p=0$ quotient and the Virasoro dressing calculation.

\section{Highest-weight BMS$_3$ blocks and conventions}\label{sec:setup}
In this section, we briefly review the BMS$_3$ algebra, primary operators and conformal blocks in the highest-weight representation,
fixing the conventions used throughout the paper.
\subsection{BMS$_3$ algebra and primary operators}
The centrally extended BMS$_3$ charge algebra is given by
\be
\ba
[L_n,L_m]&=(n-m)L_{n+m}+\frac{c_L}{12}n(n^2-1)\delta_{n+m,0}\,,\\
[L_n,M_m]&=(n-m)M_{n+m}+\frac{c_M}{12}n(n^2-1)\delta_{n+m,0}\,,\\
[M_n,M_m]&=0\,,
\ea
\ee
where $c_L,c_M$ are two central charges. The generators $L_{0,\pm1}$ and $M_{0,\pm1}$ form the global
subalgebra, for which the central terms vanish.

Primary states in the highest-weight representation are annihilated by the positive BMS$_3$ modes, but they need not be simultaneous eigenstates of $L_0$ and $M_0$. In a basis of $L_0$ eigenstates,
 primaries of the same scaling dimension may form a multiplet on
which $M_0$ acts as a nontrivial Jordan block: its diagonal entries give
their common boost weight, while the off-diagonal entries mix the
components \cite{Hao:2021urq}. A singlet has a single primary component,
which is an eigenstate of both zero modes. In this paper, the external
and exchanged primaries are set to be singlets, and a primary state
$|p\rangle\equiv|\Delta_p,\xi_p\rangle$ therefore satisfies
\be
L_{n>0}|p\rangle=M_{n>0}|p\rangle=0\,,\quad L_0|p\rangle=\Delta_p|p\rangle\,,\quad M_0|p\rangle=\xi_p|p\rangle\,.
\ee
The Verma module generated by $|p\rangle$ is spanned by products of negative modes $L_{-n},M_{-n}$ with $n\geq1$,
acting on the primary state. Its global descendants are generated by $L_{-1}$ and $M_{-1}$ alone. The bra states are defined by the Hermitian conjugation $L^\dagger_n=L_{-n},M^\dagger_n=M_{-n}$ and normalization $\langle p|p\rangle=1$.

On the plane with coordinates $(x,t)$, a finite BMS symmetry transformation is given by
\be\label{bms-transformation}
\tilde x=f(x)\,,\quad \tilde t=f'(x)t+h(x)\,.
\ee
A primary singlet $O(x,t)$ of weights $(\Delta,\xi)$ transforms according to \cite{Hao:2021urq}
\be\label{def:primary}
 O^{(f,h)}( x,t)=[f'(x)]^{\Delta}\exp\left[\xi\frac{tf''(x)+h'(x)}{f'(x)}\right]O(\tilde x,\tilde t)\,.
\ee
The corresponding infinitesimal transformations are generated by
\be
\ba\label{Commutator}
[L_n,O(x,t)]&=\left[x^{n+1}\p_x+(n+1)x^nt\p_t+(n+1)\Delta x^n+n(n+1)\xi x^{n-1}t\right]O(x,t)\,,\\
[M_n,O(x,t)]&=\left[x^{n+1}\p_t+(n+1)\xi x^n\right]O(x,t)\,.
\ea
\ee
The state--operator correspondence identifies the primary state
$|p\rangle$ with an insertion of the corresponding primary singlet
at the origin, which is $|p\rangle=O_p(0,0)|0\rangle$.

\subsection{BMS  blocks }
The two-point and three-point functions of primary operators are fixed up to normalization constants by global symmetries. Let $O_i$ be a primary operator with weights $(\Delta_i,\xi_i)$. Writing  $x_{ij}=x_i-x_j,t_{ij}=t_i-t_j$, these functions take the form \cite{Bagchi:2009ca,Bagchi:2009pe}
\be\ba
\langle O_1(x_1,t_1)O_2(x_2,t_2)\rangle&= \delta_{\Delta_1,\Delta_2}\delta_{\xi_1,\xi_2}x_{12}^{-2\Delta}e^{-2\xi\frac{t_{12}}{x_{12}}}\,,\\
\langle O_1(x_1,t_1)O_2(x_2,t_2)O_3(x_3,t_3)\rangle&=\frac{c_{123}}{x_{12}^{\Delta_{12,3}}x_{23}^{\Delta_{23,1}}x_{13}^{\Delta_{13,2}}}\exp\left[-\xi_{12,3}\frac{t_{12}}{x_{12}}-\xi_{23,1}\frac{t_{23}}{x_{23}}-\xi_{13,2}\frac{t_{13}}{x_{13}}\right]\,,
\ea
\ee
where  $\Delta_{ij,k}\equiv\Delta_i+\Delta_j-\Delta_k,\xi_{ij,k}\equiv\xi_i+\xi_j-\xi_k$. Paired primaries have unit two-point
normalization and the constant $c_{123}$ is then the OPE coefficient.

The coordinate dependence of a four-point function, after removing an external kinematic factor,
is through the two BMS cross ratios defined by
\be
x=\frac{x_{12}x_{34}}{x_{13}x_{24}}\,,\quad \frac{t}{x}=\frac{t_{12}}{x_{12}}+\frac{t_{34}}{x_{34}}-\frac{t_{13}}{x_{13}}-\frac{t_{24}}{x_{24}}\,.
\ee
We use global symmetry to fix the location of three points  and consider the following four point function
\be
G(x,t)\equiv\langle O_1(\infty,0)O_2(1,0)O_3(x,t)O_4(0,0)\rangle\,.
\ee
 Let $\Pi_p$ be the projector onto the complete module of primary $O_p$. The BMS block $\mathcal{F}_p$ is then defined as its contribution with the OPE coefficients and external factor removed
\be\ba\label{def:Fp}
\langle O_1(\infty,0)O_2(1,0)\Pi_pO_3(x,t)O_4(0,0)\rangle=c_{12p}c_{p34}x^{-\Delta_3-\Delta_4}\exp\left[-(\xi_3+\xi_4)\frac{t}{x}\right]\mathcal{F}_p(x,t)\,.
\ea
\ee
Summing over the exchanged primaries gives the complete four-point function
\be
G(x,t)=x^{-\Delta_3-\Delta_4}\exp\left[-(\xi_3+\xi_4)\frac{t}{x}\right]\sum_pc_{12p}c_{p34}\mathcal{F}_p(x,t)\,.
\ee
 In the OPE limit $x,t\to0$ with $t/x$ fixed, we have
\be\label{OPE}
\mathcal{F}_p(x,t)\sim x^{\Delta_{p} }\exp\left(\xi_p \frac{t}{x}\right)\,.
\ee
The projector $\Pi_p$ is independent of the descendant basis and can be evaluated using the Gram matrix. Let $|p;N,I\rangle$ denote a descendant state of primary $|\Delta_p,\xi_p\rangle$ of level $N$ with $I$ labeling the states at that level. It is easy to show that states with different levels are orthogonal to each other. Consequently, 
the Gram matrix at level $N$ is then given by
\be
(\mathcal{G}^p_N)_{IJ}=\langle p;N,I|p;N,J\rangle\,.
\ee
For generic weights this matrix is invertible, and the module projector is
\be\label{Pi}
\Pi_p=\sum_{N,I,J}| p;N,I\rangle (\mathcal G^p_N)_{IJ}^{-1}\langle  p;N,J|\,.
\ee
If null states are present, the construction instead requires the
Gram matrix to be evaluated within a nondegenerate quotient module.

With the normalization in \eqref{def:Fp}, we define the left and right three-point vectors by
 
\be\ba\label{def:lr-N}
(\textbf{l}_N)_I&=\frac{\langle O_1(\infty,0)O_2(1,0)|p;N,I\rangle}{c_{12p}}\,,\\
(\textbf{r}_N)_I&=\frac{x^{\Delta_3+\Delta_4}e^{(\xi_3+\xi_4)t/x}}{c_{p34}}\langle p;N,I|O_3(x,t)O_4(0,0)\rangle\,.
\ea
\ee
They are  vectors in the chosen descendant basis, with all dependence on $x,t$ carried by $\textbf{r}_N$.  
Substituting \eqref{Pi} into \eqref{def:Fp} yields
\be\label{Fp:lGr}
\mathcal{F}_{p}(x,t)=\sum_{N\geq0}(\textbf{l}_N)^T(\mathcal{G}^p_N)^{-1}\textbf{r}_N\,.
\ee
As we will see later, the descendant three-point matrix elements can be determined by the primary three-point functions through the Ward identities.

We will be considering the light BMS block obtained by taking   $c_M\to\infty$ while keeping $c_L$ and all external and exchanged
weights fixed. For generic non-vacuum exchange with $\xi_p\neq0$, the block has the expansion
\be
\mathcal F_{p}=\mathcal{F}_{p}^{(0)}+\frac{1}{c_M}\mathcal{F}_{p}^{(1)}+O(c_M^{-2})\,.
\ee
 At leading order, only the global descendants contribute
\cite{Bagchi:2016geg,Bagchi:2017cpu}. Thus, $\mathcal{F}_p^{(0)}$ is the global block, obtained  by restricting the projector to the states $L_{-1}^aM_{-1}^{N-a}|p\rangle$, $0\leq a\leq N$. For $\xi_p=0$, the global Gram matrix is already singular at level one, so a separate treatment is needed.  The non-vacuum quotient and evaluation of linearized correction to the conformal block when $\xi_p=0$ are discussed in Appendix \ref{sec:xip=0}.

For generic external weights, we define $\Delta_{ij}=\Delta_i-\Delta_j,\xi_{ij}=\xi_i-\xi_j$ and
\be\label{def:sx}
s(x)=\sqrt{1-\left(1+\frac{\xi_{12}\xi_{34}}{\xi_p^2}\right)x+\frac{(\xi_{12}+\xi_{34})^2}{4\xi_p^2}x^2}\,.
\ee
The leading global block 
in the   normalization \eqref{OPE} is determined by the global Casimir equations and takes the following form
\cite{Chen:2022cpx}. \footnote{
Relative to Eq.~(5.11) of \cite{Chen:2022cpx}, we identify
$k_{\mathrm{there}}=-t/x$ and impose the unit primary OPE
coefficient in \eqref{OPE}, replacing the prefactor
$\xi_p^{2\Delta_p-3}$ by $\xi_p^{2\Delta_p-2}$.
We also correct the printed sign of the term linear in $x$
in the bracket of the third line of \eqref{F0}:
the term is
$-\tfrac12(2\xi_p^2-\xi_{12}^2-\xi_{34}^2)x$.
This sign is fixed by the global Ward identities and agrees
with a direct level-one projector calculation.
}
\be\ba\label{F0}
\mathcal{F}^{(0)}_p(x,t)=&\frac{2^{\Delta_p-1}\xi_p^{2\Delta_p-2}}{s(x)}x^{\Delta_p}(\xi_{p1,2}\xi_{p4,3})^{-\frac{1}{2}(\Delta_{12}-\Delta_{34})}(\xi_{p1,2}\xi_{p3,4})^{-\frac{1}{2}(\Delta_{12}+\Delta_{34})}\\
&\times\left[\xi_p^2-\frac{1}{2}(\xi_p^2+\xi_{12}\xi_{34})x+\xi_p^2s(x)\right]^{1-\Delta_p}\\
&\times\left[\xi_p^2-\xi_{12}\xi_{34}-\frac{1}{2}(2\xi_p^2-\xi_{12}^2-\xi_{34}^2)x+(\xi_{12}-\xi_{34})\xi_ps(x)\right]^{\frac{1}{2}(\Delta_{12}-\Delta_{34})}
\\&\times\left[\xi_p^2+\xi_{12}\xi_{34}-\frac{1}{2}(\xi_{12}+\xi_{34})^2x+(\xi_{12}+\xi_{34})\xi_ps(x)\right]^{\frac{1}{2}(\Delta_{12}+\Delta_{34})}\\
&\times\exp\left[\frac{t}{x(1-x)}\left(\xi_ps(x)-\frac{1}{2}(\xi_{12}-\xi_{34})x\right)\right]\,.
\ea\ee
When the operators within each external pair are identical such that $O_1=O_2,O_3=O_4$, the weight differences vanish and $s(x)=\sqrt{1-x}\equiv s_x$. The global block reduces to
\be\label{F0-iden}
\mathcal{F}_p^{(0)}(x,t)\big|_{O_1=O_2,O_3=O_4}=4^{\Delta_p-1}\frac{(1+s_x)^{2-\Delta_p}(1-s_x)^{\Delta_p}}{s_x}\exp\left[\frac{\xi_pt}{s_x(1-s_x^2)}\right]\,.
\ee
In the following, we will compute the first order correction term $\mathcal{F}^{(1)}_{p}$ with $\xi_p\neq0$.

\section{Descendant truncation at large $c_M$}\label{sec:degree}

In this section, we prove, at every fixed descendant level $N$, that the coefficient of $c_M^{-1}$ receives
contributions only from the global sector and from states containing one non-global generator. In the Virasoro case, an analogous truncation is justified by the large-$c$ scaling of descendant norms  \cite{Bombini:2018jrg}.

\subsection{The PBW basis and degree decomposition}

According to the Poincar\'e-Birkhoff-Witt (PBW) theorem, ordered products of the negative modes
$L_{-n}$ and $M_{-n}$, with $n>0$, acting on $|p\rangle$
form a basis of the Verma module. A general basis state of level $N$ is given by
\be\label{PBW}
L_{-\lambda_1}\cdots L_{-\lambda_j}M_{-\mu_1}\cdots M_{-\mu_k}| p\rangle\,,\quad N=\sum_{a=1}^j\lambda_a+\sum_{b=1}^k\mu_b
\ee
with
\be
\lambda_1\geq\lambda_2\geq\cdots\geq\lambda_j\,,\quad \mu_1\geq\mu_2\geq\cdots\geq\mu_k\,.
\ee
The ordered monomial in  \eqref{PBW} is called a PBW word and its degree is the number of non-global modes.
At fixed level $N$, we let $E_{N,s}$ be the vector space spanned by PBW states of degree $s$. Then the Verma module $\mathcal V_p$ can be decomposed as follows
\be
\mathcal{V}_p=\bigoplus_{N\geq0}\mathcal{V}_{p,N}\,,\quad \mathcal{V}_{p,N}=\bigoplus_{s=0}^{[\frac{N}{2}]}E_{N,s}\,.
\ee
This should be understood as a vector space decomposition instead of an orthogonal decomposition. In particular, states with different degrees can have nontrivial inner product.
The level $N$ block of the Gram matrix $\mathcal{G}^p_N$ can be further organized into blocks of inner product between vector spaces of different degrees. More explicitly, ordering the level-N basis by increasing degree, the matrix $\mathcal{G}^p_N$ can be written as follows
\be\label{Gram-degree}
\mathcal{G}^p_N=\left(\begin{array}{cccc}
  A^{0,0}_N   & A^{0,1}_N &\cdots &A^{0,[\frac{N}{2}]}_N\\
  A^{1,0}_N   & A_N^{1,1}&\cdots &A_N^{1,[\frac{N}{2}]}\\
 \vdots & \vdots && \vdots \\
   A^{[\frac{N}{2}],0}_N   & A_N^{[\frac{N}{2}],1}&\cdots &A_N^{[\frac{N}{2}],[\frac{N}{2}]}
\end{array}\right)\,,
\ee
where
\be
A^{i,j}_N:=\langle E_{N,i}|E_{N,j}\rangle\,
\ee
denotes the block formed by the inner product between states in $E_{N,i}$ and $E_{N,j}$ respectively.

\subsection{Nondegeneracy of the diagonal Gram blocks}
The factor $c_M$ in the Gram matrix can arise only from the central part of the commutator $[L_n,M_{-n}]$ with $n\geq2$. Each such contraction consumes one non-global generator on each side of the pairing. It then follows 
that $A^{i,j}_N=O(c_M^{\text{min}(i,j)})$. In particular, the diagonal block satisfies
\be
A^{i,i}_N=c_M^i\Gamma_{N,i}+O(c_M^{i-1})\,.
\ee
In the following, we will show that $\Gamma_{N,i}$ is nondegenerate. Define
\be\label{def:dimension}
d_{N,s}:=\text{dim}~E_{N,s}\,,
\ee
and let $\text{deg}_{c_M}f$ denote the degree of $f$ as a polynomial in $c_M$.  Using the scaling property of each block $A_N^{i,j}$ we have discussed above, it is not difficult to show that 
\be
\det\mathcal{G}_N^p=c_M^{\sum_{s=0}^{[N/2]}sd_{N,s}}\left(\prod_{i=0}^{[N/2]}\det \Gamma_{N,i}+O(c_M^{-1})\right)\,.
\ee
As a consequence,
to prove that $\Gamma_{N,i}$ is nondegenerate for all $i$, it is equivalent to showing  
\be\label{goal}
\text{deg}_{c_M}\det \mathcal{G}^p_N=\sum_{s=1}^{[\frac{N}{2}]} sd_{N,s}\,.
\ee
To show that \eqref{goal} is satisfied, we use the Kac determinant
formula \cite{Jiang:2017vtt,Yu:2022bcp}, which in our conventions
takes the following form
\be
\det \mathcal{G}^p_N=C_N\left[\prod_{ab\leq N,a,b\in\mathbb N^+}\chi(a,b)^{\theta_N(a,b)}\right]^2\,.
\ee
Here, $C_N$ is a normalization constant independent of $c_M$ and \footnote{The printed formula in \cite{Yu:2022bcp} has $f(N-ab-1,a)$ instead of $f(N-ab-i,a)$ in the expression for $\theta_N(a,b)$, which is a typo.}
\be\ba
\chi(a,b)&=\left(2a\xi_p+\frac{c_M}{12}a(a^2-1)\right)^bb!\,,\\
\theta_N(a,b)&=\sum_{i=0}^{N-ab}P(i)f(N-ab-i,a)
\ea
\ee
with $P(i)$ being  the number of partitions of $i$ and $f(N,a)$ defined through
\be
F_a(q)\equiv\sum_{N\geq0} f(N,a)q^N=\prod^\infty_{k\neq a}\frac{1}{1-q^k}\,.
\ee
Firstly, we note that in the large $c_M$ limit, $\chi(1,b)= O(1)$ and $\chi(a,b)=O(c_M^b)$ for $a>1$. Therefore, 
\be
\text{deg}_{c_M}\det \mathcal{G}^p_N=\sum_{ab\leq N,a\geq2,b\geq1}2b\theta_N(a,b)\,.
\ee
It then remains to show
\be\label{goal2}
\sum_{s=1}^{[\frac{N}{2}]} sd_{N,s}=\sum_{ab\leq N,a\geq2,b\geq1}2b\theta_N(a,b)\,,
\ee
which proves \eqref{goal}.
To evaluate the left hand side, we introduce a generating function that records both the level $N$ and the degree of descendant state
\be\label{Gen:Ns}
Z(q,y)=\sum_{N\geq0} \sum_{s=0}^{[\frac{N}{2}]}d_{N,s}q^Ny^s\,.
\ee
A generic descendant can be written as
\be
\left(\prod_{n\geq1}L_{-n}^{\ell_n}\right)
\left(\prod_{n\geq1}M_{-n}^{m_n}\right)|p\rangle\,, \quad \ell_n,m_n\in\mathbb N\,.
\ee
Its level and degree are 
\be\label{value:Ns}
N=\sum_{n\geq1}n(\ell_n+m_n)\,,\quad s=\sum_{n\geq2}(\ell_n+m_n)\,.
\ee
Plugging \eqref{value:Ns} into \eqref{Gen:Ns} gives
\be\ba
Z(q,y)&=\sum_{\{\ell_n,m_n\}}q^{\sum_{n\geq1}n(\ell_n+m_n)}y^{\sum_{n\geq2}(\ell_n+m_n)}\\
&=\frac{1}{(1-q)^2}\prod_{n=2}^\infty\frac{1}{(1-yq^n)^2}\,.
\ea
\ee
As a result, we can obtain the generating function of the left hand side of \eqref{goal2}
\be\label{GF:degree}
\sum_{N\geq0}\left(\sum_ssd_{N,s}\right)q^N=y\frac{\p}{\p y}Z(q,y)\big|_{y=1}=2\left(\prod_{n=1}^\infty\frac{1}{1-q^n}\right)^2\sum_{a=2}^\infty\frac{q^a}{1-q^a}\,.
\ee
Now we consider the right hand side of \eqref{goal2}. 
The generating function of $\theta_N(a,b)$ is given by
\be\ba\label{Theta}
\Theta(a,b)&\equiv\sum_{N\geq0}\theta_N(a,b)q^N=\sum_{N\geq ab}\left[\sum_{i=0}^{N-ab}P(i)f(N-ab-i,a)\right]q^N
\\&=q^{ab}\sum_{M\geq0}\left[\sum_{i=0}^MP(i)f(M-i,a)\right]q^M\,.
\ea
\ee
Since $P(i)$ denotes the number of partitions of $i$, it satisfies
\be
\sum_{i=0}^\infty P(i)q^i=\prod_{n=1}^\infty\frac{1}{1-q^n}\,.
\ee
Using Cauchy product, \eqref{Theta} then becomes
\be
\Theta(a,b)=q^{ab}\sum_{i=0}^\infty P(i)q^i\sum_{j=0}^\infty f(j,a)q^j=q^{ab}\prod_{n=1}^\infty\frac{1}{1-q^n}F_a(q)=q^{ab}(1-q^a)\left(\prod_{n=1}^\infty\frac{1}{1-q^n}\right)^2\,.
\ee
As a consequence,
\be
\ba
{}&\sum_{N\geq0}\left[\sum_{ab\leq N,a\geq2,b\geq1}2b\theta_N(a,b)\right]q^N=2\left(\prod_{n=1}^\infty\frac{1}{1-q^n}\right)^2\sum_{a\geq2}\sum_{b\geq1}bq^{ab}(1-q^a)\\
=&2\left(\prod_{n=1}^\infty\frac{1}{1-q^n}\right)^2\sum_{a=2}^\infty\frac{q^a}{1-q^a}\,,
\ea
\ee
which agrees precisely with \eqref{GF:degree} for arbitrary $q$. Therefore, \eqref{goal2} holds and we have proved \eqref{goal}, which proves that the leading coefficient of the diagonal block  in the large $c_M$ expansion is nondegenerate. 

\subsection{Orthogonalization and descendant truncation}
We perform block Gaussian elimination to make the Gram matrix block diagonal in increasing order of degree $s$. Let $\tilde A_N^{i,i}$ be the block obtained after orthogonalizing $E_{N,i}$ against $E_{N,r}$ for all $r<i$.
The first nontrivial block is the Schur
complement
\be
\tilde A_N^{1,1}=A_N^{1,1}-A_N^{1,0}(A_N^{0,0})^{-1}A_N^{0,1}\,.
\ee
Suppose that the preceding changes of basis have made each sector of
degree $0,\ldots,r-1$ orthogonal to every sector of a different degree.
We next make all sectors of degree $i>r$ orthogonal to the sector of
degree $r$. This subtracts a Schur-complement term of order $c_M^r$
from each Gram block $A_N^{i,j}$ with $i,j>r$. It therefore preserves the leading diagonal terms of degrees greater than $r$, as well as the scaling bounds for the remaining off-diagonal blocks. Starting with the invertible global block, induction shows that every subtraction from the degree-$i$ diagonal block is at most $O(c_M^{i-1})$. Alternatively, the Gaussian elimination amounts to the following transformation
\be
\tilde{\mathcal G}_N^p=\mathcal T_N^\dagger{\mathcal G}_N^p\mathcal T_N=\left(\begin{array}{cccc}
    A^{0,0}_N  & && \\
     & \tilde A^{1,1}_N &&\\
     && \cdots&\\
     &&&\tilde A_N^{[\frac{N}{2}],[\frac{N}{2}]}
\end{array}\right)
\ee
The BMS$_3$ block then becomes
\be
\mathcal F_{ p}(x,t)=\sum_{N\geq0}(\tilde{\mathbf l}_N)^T(\tilde{\mathcal{G}}^p_N)^{-1}\tilde{\textbf{r}}_N
\ee
with $\tilde{\textbf{l}}_N=\mathcal{T}_N^T\textbf{l}_N,\tilde{\textbf{r}}_N=\mathcal{T}^\dagger_N\textbf{r}_N$.
For each diagonal block, the preceding elimination gives
\be
\tilde A_N^{ii}=A_N^{ii}-B_N^{ii}\,,
\ee
where $B_N^{ii}$ is the sum of the successive Schur subtractions
from degrees $r<i$. Since each subtraction is $O(c_M^r)$,
we have $B_N^{ii}=O(c_M^{i-1})$, so the leading term of the diagonal block is unchanged.
Consequently, this suggests
\be
\tilde A_N^{ii}=c_M^i\Gamma_{N,i}+O(c_M^{i-1})\,.
\ee
Since $\Gamma_{N,i}$ is nondegenerate, we have
\be
(\tilde A_N^{ii})^{-1}=c_M^{-i}\Gamma_{N,i}^{-1}+O(c_M^{-i-1})\,.
\ee
To determine the contribution of each sector to the block, we also
need the scaling of the transformed three-point vectors.
The original vectors $\mathbf l_N,\mathbf r_N$ are $O(1)$ because
the primary Ward identities contain no central term and all weights
are fixed in the large $c_M$ limit.
At the step that makes the sectors of degree $i>r$ orthogonal to
the degree-$r$ sector, the required coefficients are obtained by
multiplying the inverse degree-$r$ Gram block $(\tilde A^{rr}_N)^{-1}$, which is
$O(c_M^{-r})$, by the mixed  Gram blocks at that stage, which are 
$O(c_M^r)$. These coefficients are therefore $O(1)$.
Since there are finitely many steps at fixed $N$, we have
$\mathcal T_N=O(1)$, and hence
$\tilde{\mathbf l}_N,\tilde{\mathbf r}_N=O(1)$. As a result, the new three point vectors $\tilde{\mathbf{l}}_N$ and $\tilde{\mathbf{r}}_N$ are $O(1)$. Moreover, since $A_N^{0i}$ is independent of $c_M$, the transition matrix is actually independent of $c_M$ when restricted to $E_{N,0}\oplus E_{N,1}$.
This proves that to compute the $1/c_M$ correction, it is enough to keep only the first two diagonal blocks.

\section{Evaluation of the linearized correction}\label{sec:intrinsic}
In this section, 
we compute the contribution to the linearized correction $\mathcal{F}_p^{(1)}$ at fixed level $N$  and then take a sum over $N$. We begin by constructing a suitable basis for $E_{N,0}\oplus E_{N,1}$ such that the Gram matrix restricted to this subspace takes the block diagonal form. Once the basis has been determined, we compute the three elements on the right hand side of \eqref{Fp:lGr}: the two three-point vectors and the corresponding Gram matrix.

\subsection{Descendant basis and Gram matrices}

A basis for the global sector $E_{N,0}$ can be chosen to be PBW states. More explicitly, we define a vector valued operator $D_N$ whose components are
\be
(D_N)_a=L_{-1}^aM_{-1}^{N-a}\,,\quad a=0,\cdots,N\,.
\ee
The states $D_N|p\rangle$ span $E_{N,0}$ and their Gram matrix is given by \footnote{Here $D_N$ is regarded as an ordered list of operators, with $\dagger$ acting on each component. Products such as $D_N^\dagger D_N$ denote matrices of ordered operator products with the first list labeling the rows and the second labeling the columns.}
\be\label{global-Gram}
g_N(\Delta_p,\xi_p)=\langle p|D_{N}^\dagger D_{N}| p\rangle\,.
\ee
Moreover,
we define two vectors of normalized three-point functions $\mathbf l_N^{\text{gl}}$,$\mathbf r_N^{\text{gl}}$ for each $N$ as
\be\ba\label{def:lr-gl}
{}&\mathbf{l}_N^{\text{gl}}=\frac{\langle O_1(\infty,0)O_2(1,0)D_{N}|p\rangle}{c_{12p} }\,,\quad \mathbf{r}_N^{\text{gl}}=\frac{x^{\Delta_3+\Delta_4}e^{(\xi_3+\xi_4)t/x}}{c_{p34} }\langle  p|D^\dagger_NO_3(x,t)O_4(0,0)\rangle\,.
\ea
\ee
Note that $\mathbf{r}_N^{\text{gl}}$ depends on $x$ and $t$. 
Then the global BMS block can be written as
\be\label{F0:sum}
\mathcal{F}_{p}^{(0)}=\sum_{N\geq0}\left[(\mathbf l_N^{\text{gl}})^T(g_N)^{-1} \mathbf r_N^{\text{gl}}\right]\,.
\ee
To compute the correction at order $c_M^{-1}$, we extend the global
basis to a basis of $E_{N,0}\oplus E_{N,1}$, choosing the additional
states to be orthogonal to $E_{N,0}$. To this end, for each $m\geq2$
we construct two quasi-primary states $|U_m\rangle$ and
$|V_m\rangle$ satisfying
\be\label{condition}
L_1|U_m\rangle=M_1|U_m\rangle=L_1|V_m\rangle=M_1|V_m\rangle=0\,,\quad M_0|U_m\rangle=\xi_p|U_m\rangle\,,\quad (M_0-\xi_p)|V_m\rangle=m|U_m\rangle\,.
\ee
Besides, both states have $L_0$ eigenvalue $\Delta_p+m$.
Their normalization is fixed by
\be
|U_m\rangle=M_{-m}|p\rangle+\cdots\,,\quad |V_m\rangle=L_{-m}|p\rangle+\cdots\,,
\ee
where the omitted terms are linear combinations of level-$m$
descendants constructed from $L_{-n}$ and $M_{-n}$
with $1\leq n<m$. In particular, the state $|U_m\rangle$ can be constructed using only the lowering modes $M_{-n}$ as follows
\be\label{Um-fm}
|U_m\rangle=f_m(\{M_{-n}\};\xi_p)|p\rangle\,,\quad f_m=\sum_{k=1}^ma_{m,k}M_{-k}M_{-1}^{m-k}\,,\quad a_{m,m}=1\,.
\ee
Appendix B determines $a_{m,k}$ by solving \eqref{condition} and 
 constructs $V_m$ explicitly as well. Having obtained $|U_m\rangle$ and $|V_m\rangle$, we can construct level-$N$ states using
\be
D_{q}|U_m\rangle\,,\quad D_{q}|V_m\rangle
\ee
with $m=2,\cdots,N$ and  $q=N-m$. There are $2(q+1)$ such states for each $m$. Together with the $N+1$ states in $D_N|p\rangle$, the total number of constructed states is 
\be
N+1+2\sum_{m=2}^N(N-m+1)=N^2+1=\operatorname{dim}(E_{N,0}\oplus E_{N,1})\,.
\ee
The nondegeneracy of their Gram matrix, which will be established below, proves linear independence of these states. Combined with the dimension count, this shows that these states and the global descendants form a basis of $E_{N,0}\oplus E_{N,1}$.

The quasi-primary conditions \eqref{condition} first imply 
\be
\langle p|D_{N}^\dagger D_{q}|U_m\rangle=\langle p|D_{N}^\dagger D_{q}|V_m\rangle=0\,.
\ee
Indeed, after commuting the positive global generators to the right, the excess positive level must act on $|U_m\rangle$ or $|V_m\rangle$, where it vanishes. 
 The same argument shows that the global descendants built
on different values of $m$ are orthogonal. Therefore, only inner products at the same $m$ remain to be calculated. For fixed $q,m$, denote this  Gram block by
\be
G_{q,m}\equiv\left(\begin{array}{cc}
G_{q,m}^{(UU)}   &  G_{q,m}^{(UV)} \\
   G_{q,m}^{(VU)}   & G_{q,m}^{(VV)}  
\end{array}\right)
\ee
with
\be
G_{q,m}^{(XY)}=\langle X_m|D^\dagger_{q}D_{q}|Y_m\rangle\,,\quad X,Y=U,V\,.
\ee
Its entries are determined by the inner products of $|U_m\rangle,|V_m\rangle$ and the action of the global algebra.
Since $|U_m\rangle$ contains only  mutually commuting $M$ generators, we have $\langle U_m|U_m\rangle=0$. Equations \eqref{Vm0} and \eqref{Vm} show that $|V_m\rangle$ contains at most one $L$ generator. Every term in $|U_m\rangle$ other than $M_{-m}|p\rangle$ contains at least two $M$ generators which leads to a vanishing overlap with $|V_m\rangle$. Therefore,
the only nontrivial contribution to $\langle U_m|V_m\rangle$ comes from 
\be\label{inner:UV}
\mathcal N_m\equiv\langle U_m|V_m\rangle=\langle p|M_mL_{-m}| p\rangle=\frac{c_M}{12}m(m^2-1)+2m\xi_p\,.      
\ee
To identify the term proportional to $c_M$ in $\langle V_m|V_m\rangle$,
we use \eqref{Vm0} and \eqref{Vm}. Each monomial contains at most one
$L$ generator, placed to the left of all $M$ generators.
Since removing an $M$ generator requires a commutator with an $L$
generator, a nonzero pairing can contain at most two $M$ generators in total.
The pure-$M$ terms contain at least two $M$ generators, since
$a_{m,m}=1$ is independent of $\xi_p$, and therefore have zero overlap
with every term in $V_m$.
A term containing one $L$ and two $M$ generators can only pair with
$L_{-m}|p\rangle$. Commuting $L_m$ past its $L_{-k}$ leaves
$L_{m-k}$ acting on two $M$ generators of total level $m-k$.
Each commutator then produces a positive $M$ mode, which commutes
with the remaining $M$ mode and annihilates $|p\rangle$.
Thus only terms containing one $L$ and at most one $M$ generator
need be retained. Reading them from \eqref{Vm0} gives
\be
L_{-m}| p\rangle\,,\quad a_{m,m-1}L_{1-m}M_{-1}| p\rangle\,,\quad a_{m,m-1}L_{-1}M_{1-m} |p\rangle\,.
\ee
where $a_{m,m-1}=-\frac{m+1}{2\xi_p}$ for $m>2$ according to \eqref{coe:a}. 
Let $\mathcal{D}_m\equiv\langle V_m|V_m\rangle$. Direct computation shows
\be\label{inner:VV}
\lim_{c_M\to\infty}\frac{\mathcal D_m}{c_M}=\frac{m(m-1)(m-2)}{12}[4\xi_pa^2_{m,m-1}+2(m+1)a_{m,m-1}]=0\,.
\ee
Thus, $\langle V_m|V_m\rangle=O(1)$.
Since $|U_m\rangle$ and $|V_m\rangle$ are annihilated by $L_1,M_1$, we have
\be\label{inner:XDDY}
\langle X_m|D^\dagger_{q}D_{q}|Y_m\rangle=\langle X_m|g_q(L_0,M_0)|Y_m\rangle\,,
\ee
where the matrix $g_q$ is defined in \eqref{global-Gram} with $N$ replaced by $q$ and weights $(\Delta_p,\xi_p)$ replaced by the commuting generators $(L_0,M_0)$. The action of $M_0$ in \eqref{condition} gives
\be
M_0^r|U_m\rangle=\xi_p^r|U_m\rangle\,,\quad M_0^r|V_m\rangle=\xi_p^r|V_m\rangle+mr\xi_p^{r-1}|U_m\rangle
\ee
for every nonnegative integer $r$. Since  $L_0=\Delta_p+m$ on both $|U_m\rangle$ and $|V_m\rangle$, we have
\be\ba\label{g-UV}
g_q(L_0,M_0) |U_m\rangle&=g_q(\Delta_p+m,\xi_p) |U_m\rangle\,,\\
g_q(L_0,M_0)|V_m\rangle&=g_q(\Delta_p+m,\xi_p) |V_m\rangle+m\p_{\xi_p}g_q(\Delta_p+m,\xi_p) |U_m\rangle\,.
\ea
\ee
Combining \eqref{inner:XDDY} with \eqref{g-UV} yields
\be
 G_{q,m}= \left(\begin{array}{cc}
  0   & \mathcal N_mg_q(\Delta_p+m,\xi_p) \\
 \mathcal N_mg_q(\Delta_p+m,\xi_p)  &  \mathcal D_mg_q(\Delta_p+m,\xi_p)+m\mathcal N_m\p_{\xi_p}g_q(\Delta_p+m,\xi_p)
\end{array}\right) \,.
\ee
 For $\xi_p\neq0$, the global Gram matrix is nondegenerate, so its inverse exists.
Using \eqref{inner:UV} and \eqref{inner:VV}, we find
\be\label{Ginverse:degree1}
G_{q,m}^{-1}=\frac{12}{c_Mm(m^2-1)}\left(\begin{array}{cc}
   m\p_{\xi_p}g_q^{-1}(\Delta_p+m,\xi_p)  & g_q^{-1} (\Delta_p+m,\xi_p)\\
   g_q^{-1} (\Delta_p+m,\xi_p) & 0
\end{array}\right)+O(c_M^{-2})
\ee
in the large $c_M$ expansion. So far, organizing the basis of $E_{N,0}\oplus E_{N,1}$ into the order
\be
D_N|p\rangle\,,\quad D_q|U_m\rangle\,,\quad D_q|V_m\rangle\,,
\ee
the restricted Gram matrix takes the form
\be
\left.\mathcal G_N^p\right|_{E_{N,0}\oplus E_{N,1}}
=g_N\oplus\bigoplus_{m=2}^{N}G_{N-m,m}\,,
\ee
which is clearly nondegenerate at sufficiently large $c_M$. This proves that these states are linearly independent.
These
Gram matrices determine the inverse pairing in the block. We next compute the three-point
matrix elements in the same basis.
\subsection{Three-point vectors}
On the basis of $E_{N,0}\oplus E_{N,1}$ constructed above, we will show that the three-point vectors are determined by the
global vectors defined in \eqref{def:lr-gl} and their first derivatives with respect to $\xi_p$. The global vectors  can
be generated from the normalized primary three-point functions
\be\ba\label{def:W}
\mathcal{W}_{\Delta_p,\xi_p}(x,t)&\equiv\frac{\langle O_1(\infty,0)O_2(1,0)O_p(x,t)\rangle}{c_{12p}}=(1-x)^{-\Delta_{p2,1}}\exp\left(\frac{\xi_{p2,1}t}{1-x}\right)\,,\\
\widetilde{\mathcal{W}}_{\Delta_p,\xi_p}(x,t)&\equiv x^{\Delta_3+\Delta_4}e^{(\xi_3+\xi_4)\frac{t}{x}}\frac{\langle p|O_3(x,t)O_4(0,0)\rangle}{c_{p34}}=x^{\Delta_p}e^{\xi_pt/x}\,.
\ea
\ee
The external factor in the second expression is the one required by the block normalization in \eqref{def:Fp}. Using Ward identities, global vectors are determined by
\be\ba\label{rel:vec-W}
(\mathbf l^{\text{gl}}_N)_a&=\p_x^a\p_t^{N-a}\mathcal{W}_{\Delta_p,\xi_p}(x,t)\big|_{x=t=0}\,,\\
(\mathbf r^{\text{gl}}_N)_a&=(x^2\p_t+\xi_{34}x)^{N-a}(x^2\p_x+2xt\p_t+\Delta_{34}x+\xi_{34}t)^a\widetilde{\mathcal W}_{\Delta_p,\xi_p}(x,t)\,.
\ea\ee
Appendix \ref{app:3pv} derives these differential relations and gives the resulting components explicitly.

We now compute the three-point matrix elements of $U_m,V_m$.
For an ordered pair of external operators $O_i,O_j$,  define their couplings with $U_m,V_m$ by
\be\label{def:lambda}
\lambda^X_{ij,m}=\frac{\langle O_i(\infty,0)O_j(1,0)|X_m\rangle}{c_{ijp}}\,,\quad X=U,V\,.
\ee
The Ward identity then gives
\be\label{Ward-3pt-M}
\frac{\langle O_i(\infty,0)O_j(1,0)M_{-k}M_{-1}^{m-k}| p\rangle}{c_{ijp}}=(\xi_p-\xi_{ij})^{m-k}(\xi_p+k\xi_j-\xi_i)\,.
\ee
Substituting the polynomial $f_m$ introduced in \eqref{Um-fm} and its coefficients from \eqref{coe:a}, we obtain
\be\ba\label{lamU}
\lambda^U_{ij,m}&=\sum_{k=1}^ma_{m,k}(\xi_p-\xi_{ij})^{m-k}(\xi_p+k\xi_j-\xi_i)\\
&=\frac{1}{(2\xi_p)^{m+1}}\left[(\xi_p+\xi_{ij})^m(\xi_p^2-\xi_i^2+\xi_j^2+2m\xi_p\xi_j)\right.
\\&\quad\quad\quad~~~~\left.+(-1)^m(\xi_p-\xi_{ij})^m(\xi_p^2+\xi_i^2-\xi_j^2+2m\xi_p\xi_i)\right]\,.
\ea
\ee
The coupling of $V_m$ follows from \eqref{Vm} with $V_m^{(0)}$ given by \eqref{Vm0}.
From \eqref{Ward-3pt-M}, we find that each $M_{-k}$ factor contributes to the normalized three point function by the factor $\xi_p+k\xi_j-\xi_i$. Similarly, it can be shown that the contribution from one $L_{-k}$ mode is given by the factor $\Delta_p+m-k+k\Delta_j-\Delta_i$. More precisely, for any state $|\chi\rangle$ satisfying
$L_0|\chi\rangle=(\Delta_p+m-k)|\chi\rangle$, the Ward identity gives
\be
\frac{\langle O_i(\infty,0)O_j(1,0)L_{-k}|\chi\rangle}{c_{ijp}}
=
(\Delta_p+m-k+k\Delta_j-\Delta_i)
\frac{\langle O_i(\infty,0)O_j(1,0)|\chi\rangle}{c_{ijp}}\,.
\ee
As a result, we have
\be\ba\label{lamV}
\lambda_{ij,m}^V&=\left[\sum_{n=1}^m[\Delta_p+m-n+n\Delta_j-\Delta_i]\frac{\p f_m}{\p M_{-n}}+(\Delta_p+m-1)\frac{\p f_m}{\p\xi_p}\right]\Big|_{\{M_{-n}\}=\{\xi_p+n\xi_j-\xi_i\}}\\
&=\left[(\Delta_p+m-1)\p_{\xi_p}+(\Delta_i-1)\p_{\xi_i}+(\Delta_j-1)\p_{\xi_j}\right]\lambda_{ij,m}^U\,.
\ea
\ee
In the first line, the derivative $\p_{\xi_p}f_m$ is taken at fixed formal variables $M_{-n}$ before imposing the displayed
substitution. 

It remains to include the global descendants of $U_m,V_m$. The corresponding left three-point vectors are defined  by
\be
\mathbf{l}_{q,m}^X=\frac{\langle O_1(\infty,0)O_2(1,0)D_q|X_m\rangle}{c_{12p}}\,,\quad X=U,V\,.
\ee
Since $|U_m\rangle$ is quasi-primary of weights $(\Delta_p+m,\xi_p)$, the three point function $\langle O_1O_2U_m\rangle$ should be fixed by global Ward identities up to an overall constant. Comparing \eqref{def:W} with \eqref{def:lambda} and using the fact that $\mathcal{W}_{\Delta_p,\xi_p}(0,0)=1$, we have
\be
\frac{\langle O_1(\infty,0)O_2(1,0)U_m(x,t)\rangle}{c_{12p}}=\lambda^U_{12,m}\mathcal{W}_{\Delta_p+m,\xi_p}(x,t)\,.
\ee
Components of the vector $\mathbf l_{q,m}^U$ can be obtained by taking derivatives $\p_x^a\p_t^{q-a}$ with respect to coordinates in the same way as the global vector $\mathbf l_q^{\text{gl}}$ in \eqref{rel:vec-W}. It is therefore straightforward to get the relation
\be\label{leftu-gl}
\mathbf l_{q,m}^U=\lambda^U_{12,m}\mathbf l_q^{\text{gl}}(\Delta_p+m,\xi_p)\,.
\ee
To obtain $\mathbf  l_{q,m}^V$, we introduce a parameter $\varepsilon$ which is nilpotent $\varepsilon^2=0$. Then the state $|U_m\rangle+\varepsilon|V_m\rangle$ satisfies
\be
M_0(|U_m\rangle+\varepsilon|V_m\rangle)=(\xi_p+m\varepsilon)(|U_m\rangle+\varepsilon|V_m\rangle)\,.
\ee
Since both $U_m$ and $V_m$ are annihilated by $L_1,M_1$,
$|U_m\rangle+\varepsilon|V_m\rangle$ is a quasi-primary state of weights $(\Delta_p+m,\xi_p+m\varepsilon)$ whose normalized three-point function with $O_1$ and $O_2$ is given by
\be
(\lambda_{12,m}^U+\varepsilon\lambda^V_{12,m})\mathcal{W}_{\Delta_p+m,\xi_p+m\varepsilon}=\lambda^U_{12,m}\mathcal{W}_{\Delta_p+m,\xi_p}+\varepsilon[\lambda_{12,m}^V\mathcal{W}_{\Delta_p+m,\xi_p}+m\lambda^U_{12,m}\p_{\xi_p}\mathcal{W}_{\Delta_p+m,\xi_p}]\,.
\ee
From the coefficient of $\varepsilon$, we find
\be
\frac{\langle O_1(\infty,0)O_2(1,0)V_m(x,t)\rangle}{c_{12p}}=\lambda^V_{12,m}\mathcal{W}_{\Delta_p+m,\xi_p}(x,t)+m\lambda^U_{12,m}\p_{\xi_p}\mathcal{W}_{\Delta_p+m,\xi_p}(x,t)\,.
\ee
Applying $\p_x^a\p_t^{q-a}$ and then setting $x=t=0$, as before, gives
\be\label{leftv-gl}
\mathbf l_{q,m}^V=\left[\lambda^V_{12,m}\mathbf l^{\text{gl}}_q+m\lambda^U_{12,m}\p_{\xi_p}\mathbf l_q^{\text{gl}}\right](\Delta_p+m,\xi_p)\,.
\ee
A parallel discussion can be applied to the right sector.  Retaining the external normalization factor in \eqref{def:lr-N}, we define
\be
\mathbf r^X_{q,m}=\frac{x^{\Delta_3+\Delta_4}e^{(\xi_3+\xi_4)t/x}}{c_{p34} }\langle X_m|D_q^\dagger O_3(x,t)O_4(0,0)\rangle
\ee
Global covariance fixes the three point functions containing $U_m$ and $V_m$ to be
\be
\frac{x^{\Delta_3+\Delta_4}e^{(\xi_3+\xi_4)t/x}}{c_{p34}}
\langle U_m|O_3(x,t)O_4(0,0)\rangle
=\lambda_{43,m}^{U}\widetilde{\mathcal W}_{\Delta_p+m,\xi_p}(x,t)\,.
\ee
Applying differential operators as in the second line of \eqref{rel:vec-W} gives
\be
\ba\label{rightuv-gl}
\mathbf r^U_{q,m}=\lambda_{43,m}^U\mathbf r_q^{\text{gl}}(\Delta_p+m,\xi_p)\,,\quad \mathbf r_{q,m}^V=\left[\lambda^V_{43,m}\mathbf r^{\text{gl}}_q+m\lambda^U_{43,m}\p_{\xi_p}\mathbf r_q^{\text{gl}}\right](\Delta_p+m,\xi_p)\,.
\ea
\ee
Therefore, we have established
the reduction stated at the beginning of the subsection: all required components follow from the
global vectors, their first $\xi_p$ derivatives, and the two couplings for each external pair.

\subsection{Correction at order $c_M^{-1}$}
We now combine the Gram inverse with the three-point vectors at fixed level and sum over $N$ in the end. The degree-zero contribution is the global block given by \eqref{F0:sum}.
The global Gram matrix and its three-point vectors are independent of $c_M$, so this contribution has no subleading correction.

For degree one states, the orthogonality established above allows us to compute each pair $(m,q)$ separately. Substitution of the Gram inverse \eqref{Ginverse:degree1} and the left and right vectors gives
\be\ba\label{F1:mq}
{}&((\mathbf l_{q,m}^U)^T,(\mathbf l_{q,m}^V)^T)G_{q,m}^{-1}\left(\begin{array}{c}
    \mathbf  r_{q,m}^U \\
  \mathbf    r_{q,m}^V 
\end{array}\right)
\\=&\frac{12}{c_Mm(m^2-1)}\left[m(\mathbf l_{q,m}^U)^T\p_{\xi_p}g_q^{-1}\mathbf r_{q,m}^U+(\mathbf l_{q,m}^U)^Tg_q^{-1}\mathbf r_{q,m}^V+(\mathbf l_{q,m}^V)^Tg_q^{-1}\mathbf r_{q,m}^U\right]+O(c_M^{-2})
\\=&\frac{12(\lambda_{12,m}^U\lambda^V_{43,m}+\lambda^V_{12,m}\lambda^U_{43,m})}{c_Mm(m^2-1)}(\mathbf l_q^{\text{gl}})^Tg_q^{-1}\mathbf r_q^{\text{gl}}+\frac{12\lambda_{12,m}^U\lambda_{43,m}^U}{c_M(m^2-1)}\p_{\xi_p}\left[(\mathbf l_q^{\text{gl}})^Tg_q^{-1}\mathbf r_q^{\text{gl}}\right]+O(c_M^{-2})\,,
\ea
\ee
where we have used \eqref{leftu-gl},\eqref{leftv-gl} and \eqref{rightuv-gl} in the last line. All global quantities in the last line are evaluated at $(\Delta_p+m,\xi_p)$.
Summing the leading term of \eqref{F1:mq} over $q\geq0$ and using \eqref{F0:sum} gives the coefficient of $c_M^{-1}$ at fixed $m$
\be
\frac{12(\lambda_{12,m}^U\lambda^V_{43,m}+\lambda^V_{12,m}\lambda^U_{43,m})}{m(m^2-1)}\mathcal{F}^{(0)}_{\Delta_p+m,\xi_p}+\frac{12{\lambda_{12,m}^U\lambda_{43,m}^U}}{m^2-1} \p_{\xi_p}\mathcal{F}^{(0)}_{\Delta_p+m,\xi_p}\,.
\ee
Plugging \eqref{lamU} and \eqref{lamV} into the above expression and summing over $m$ gives the complete $c_M^{-1}$ correction to the BMS$_3$ block 
\be\label{F1:F0}
\mathcal{F}_{ p}^{(1)}=\sum_{m=2}^\infty\frac{12}{ m(m^2-1)}\left[(\lambda_{12,m}^U\lambda^V_{43,m}+\lambda^V_{12,m}\lambda^U_{43,m})+m\lambda^U_{12,m}\lambda^U_{43,m}\p_{\xi_p}\right]\mathcal{F}^{(0)}_{\Delta_p+m,\xi_p}\,.
\ee
This expression gives the complete first-order coefficient as a
sum of global blocks. To evaluate that sum, we use the dependence of the global block and the
couplings on $m$.
Using \eqref{F0}, we have the following relation
\be\label{rel:F}
\mathcal{F}^{(0)}_{\Delta_p+m,\xi_p}=\rho^m\mathcal{F}^{(0)}_{p}\,,\quad \p_{\xi_p}\mathcal{F}^{(0)}_{\Delta_p+m,\xi_p}=\rho^m\mathcal{F}^{(0)}_{ p}\left[\p_{\xi_p}\log\mathcal{F}^{(0)}_{p}+m\p_{\xi_p}\log\rho\right]\,,
\ee
where
\be
\rho=\frac{2\xi_p^2x}{\xi_p^2-\frac{1}{2}(\xi_p^2+\xi_{12}\xi_{34})x+\xi_p^2s(x)}
\ee
with $s(x)$ defined in \eqref{def:sx}. Plugging \eqref{rel:F} into \eqref{F1:F0} gives
\be\ba\label{F1-F0-p}
\mathcal{F}_p^{(1)}=12\mathcal{F}_p^{(0)}&\sum_{m=2}^\infty\frac{\rho^m}{m(m^2-1)}\left[\lambda_{12,m}^U\lambda^V_{43,m}+\lambda^V_{12,m}\lambda^U_{43,m}\right.\\
&\left.+m\lambda^U_{12,m}\lambda^U_{43,m}\left(\p_{\xi_p}\log\mathcal{F}_p^{(0)}+m\p_{\xi_p}\log\rho\right)\right]
\ea\ee
The global block and its logarithmic derivative no longer depend on $m$. The remaining dependence lies in the prefactor as well as the couplings.  We write 
\be\ba
\lambda_{ij,m}^U&=\sum_{\sigma=\pm1}\lambda^{U,\sigma}_{ij,m}\left(\frac{\xi_i-\xi_j+\sigma\xi_p}{2\xi_p}\right)^m\,,\quad\lambda_{ij,m}^V=\sum_{\sigma=\pm1}\lambda^{V,\sigma}_{ij,m}\left(\frac{\xi_i-\xi_j+\sigma\xi_p}{2\xi_p}\right)^m\,,
\ea
\ee
where
\be\ba
\lambda^{U,\sigma}_{ij,m}&=\frac{\xi_p^2-\sigma(\xi_i^2-\xi_j^2)+m\xi_p[(1+\sigma)\xi_j+(1-\sigma)\xi_i]}{2\xi_p}\,,\\
\lambda^{V,\sigma}_{ij,m}&=\left(\frac{\xi_i-\xi_j+\sigma\xi_p}{2\xi_p}\right)^{-m}\left[(\Delta_p+m-1)\p_{\xi_p}+(\Delta_i-1)\p_{\xi_i}+(\Delta_j-1)\p_{\xi_j}\right]\\
&~\times\left[\lambda^{U,\sigma}_{ij,m}\left(\frac{\xi_i-\xi_j+\sigma\xi_p}{2\xi_p}\right)^m\right]\,.
\ea
\ee
It is easy to see that the first coefficient is linear in $m$ while the second coefficient is a polynomial in $m$ of degree at most three. Multiplying the two decompositions for the left and right pairs now gives the combinations that
occur in \eqref{F1-F0-p}:
\be\ba\label{coe-AB}
\lambda^U_{12,m}\lambda^V_{43,m}+\lambda^V_{12,m}\lambda^U_{43,m}&=\sum_{\sigma,\tau=\pm1}A^{\sigma\tau}_m\left[\frac{(\xi_1-\xi_2+\sigma\xi_p)(\xi_4-\xi_3+\tau\xi_p)}{4\xi_p^2}\right]^m\,,\\
m\lambda^U_{12,m}\lambda^U_{43,m}&=\sum_{\sigma,\tau=\pm1}B^{\sigma\tau}_m\left[\frac{(\xi_1-\xi_2+\sigma\xi_p)(\xi_4-\xi_3+\tau\xi_p)}{4\xi_p^2}\right]^m\,,
\ea\ee
where coefficients
\be\ba
A^{\sigma\tau}_m&=\lambda^{U,\sigma}_{12,m}\lambda^{V,\tau}_{43,m}+\lambda^{V,\sigma}_{12,m}\lambda^{U,\tau}_{43,m}\,,\\
B^{\sigma\tau}_m&=m\lambda^{U,\sigma}_{12,m}\lambda^{U,\tau}_{43,m}\,,
\ea\ee
are polynomials of $m$.
In each summand on the right hand side of \eqref{coe-AB}, the displayed power combines with $\rho^m$ into a single power $(\tilde\rho_{\sigma\tau})^m$ with
\be
\tilde\rho_{\sigma\tau}=\frac{(\xi_1-\xi_2+\sigma\xi_p)(\xi_4-\xi_3+\tau\xi_p)\rho}{4\xi_p^2}\,.
\ee
Consequently, \eqref{F1-F0-p} becomes
\be
\ba\label{Fp-sum}
\mathcal{F}_p^{(1)}=12\mathcal{F}_p^{(0)}\sum_{\sigma,\tau=\pm1}\sum_{m=2}^\infty\frac{(\tilde\rho_{\sigma\tau})^m}{m(m^2-1)}\left[A_m^{\sigma\tau}+(\p_{\xi_p}\log\mathcal{F}_p^{(0)})B_m^{\sigma\tau}+(\p_{\xi_p}\log\rho) mB_m^{\sigma\tau}\right]\,.
\ea
\ee
It therefore suffices to sum the prefactor and then generate its
polynomial factors in the bracket by differentiation. Defining
\be
\Phi(z)=\sum_{m=2}^\infty\frac{z^m}{m(m^2-1)}=\frac{3z}{4}-\frac{1}{2}+\left(1-\frac{z}{2}-\frac{1}{2z}\right)\log(1-z)\,,\quad \Theta_z=z\frac{d}{dz}\,,
\ee
\eqref{Fp-sum} can be written as
\be
\ba\label{F1}
{\mathcal{F}^{(1)}_{p}}=12{\mathcal{F}^{(0)}_{p}}\sum_{\sigma,\tau=\pm1}\left[A^{\sigma\tau}_{\Theta_{\tilde\rho_{\sigma\tau}}}+(\p_{\xi_p}\log\mathcal{F}^{(0)}_{p})B^{\sigma\tau}_{\Theta_{\tilde\rho_{\sigma\tau}}}+(\p_{\xi_p}\log\rho)\Theta_{\tilde\rho_{\sigma\tau}}B^{\sigma\tau}_{\Theta_{\tilde\rho_{\sigma\tau}}}\right]\Phi(\tilde\rho_{\sigma\tau})\,,
\ea
\ee
where $A_{\Theta_{\tilde\rho_{\sigma\tau}}}^{\sigma\tau}$ and $B_{\Theta_{\tilde\rho_{\sigma\tau}}}^{\sigma\tau}$
denote the differential operators obtained by replacing $m$ with
${\Theta_{\tilde\rho_{\sigma\tau}}}$ in the polynomials $A_m^{\sigma\tau}$ and $B_m^{\sigma\tau}$.
When operators are pairwise identical such that $O_1=O_2,O_3=O_4$, \eqref{F1} can be simplified dramatically and we have
\be\label{F1:pairiden}
\mathcal{F}^{(1)}_{ p}(x,t)=12\mathcal{F}_{ p}^{(0)}[P(x)+tQ'(x)]\,,
\ee
where functions $P(x),Q(x)$ are given by
\be\ba\label{functions:PQ}
Q(x)=&-2\xi_1\xi_3\left(1+\frac{1+s_x^2}{1-s_x^2}\log s_x\right)+(\xi_1+\xi_3)\xi_p\left(1+\frac{2s_x}{1-s_x^2}\log s_x\right)\\
&+\xi_p^2\left(\log\frac{2}{1+s_x}-\frac{1}{2}-\frac{s_x^2}{1-s_x^2}\log s_x\right)\,,\\
P(x)=&\left[\left(\Delta_1-\frac{1}{2}\right)\p_{\xi_1}+\left(\Delta_3-\frac{1}{2}\right)\p_{\xi_3}+(\Delta_p-1)\p_{\xi_p}\right]Q(x)\,.
\ea\ee
\section{Linearized correction from boundary gravitons}\label{sec:dressing}
The intrinsic calculation in the previous section derives the coefficient of $c_M^{-1}$ directly from the highest-weight projector. In this section, we reproduce the same $c_M^{-1}$ correction for
two pairs of identical external operators with $O_1=O_2,O_3=O_4$ using  boundary graviton fluctuations of the shadow
representation of the global block. This is motivated by the   proposal to extend the reparametrization formalism to non-vacuum Virasoro blocks by combining reparametrized
three-point functions through a shadow transform and contracting their reparametrization modes \cite{Nguyen:2022xsw}.

We first express the global BMS block as
an integral of two three-point functions joined by the two-point
function of the exchanged primary. We then vary the three-point
functions under infinitesimal BMS transformations and evaluate
the contraction of their variations using the boundary graviton
propagators in \cite{Merbis:2019wgk}. The same method is applied to the Virasoro block and reproduces the known order $c^{-1}$ correction for two identical external pairs in appendix \ref{app:virasoro-dressing}.

\subsection{The global block as a shadow integral}

Let $O_{\tilde p}$ denote the shadow of the exchanged primary with
\be
(\Delta_{\tilde p},\xi_{\tilde p})=(2-\Delta_p,-\xi_p)\,.
\ee
For external insertions at $w_a=(x_a,t_a)$ and $w_b=(x_b,t_b)$ and an internal point $w_c=(x_c,t_c)$ with $x_a>x_c>x_b$, the normalized three-point function
with a shadow insertion is defined as
\be\ba\label{shadowkernel}
K_i(w_a,w_b,w_c)&=\frac{x_{ab}^{2\Delta_i}e^{2\xi_i \frac{t_{ab}}{x_{ab}}}}{c_{i\tilde p i}}
\langle O_i(w_a)O_{\tilde p}(w_c)O_i(w_b)\rangle\\
&=\left(\frac{x_{ac}x_{cb}}{x_{ab}}\right)^{\Delta_p-2}\exp\left[\xi_p\left(\frac{t_{ac}}{x_{ac}}+\frac{t_{bc}}{x_{bc}}-\frac{t_{ab}}{x_{ab}}\right)\right]\,,
\ea
\ee
The global OPE block $\mathcal B_{iip}^{\operatorname{gl}}$ collects the   contribution of $O_p$ and its global descendants to the
normalized product of the two external operators. With the second insertion at the origin, its definition is
\be\label{OPEblock:gl}
\Pi^{\operatorname{gl}}_pO_i(w)O_i(0)|0\rangle=c_{iip}\langle O_i(w)O_i(0)\rangle\mathcal B_{iip}^{\operatorname{gl}}(w,0)|0\rangle\,,
\ee
where $\Pi^{\operatorname{gl}}_p$ is constructed as in \eqref{Pi} with only global descendants included.  Translation defines the same object for arbitrary endpoints. Its
shadow representation is \cite{Chen:2022cpx}
\be\label{block-integral}
\mathcal{B}^{\text{gl}}_{11p}(w_1,w_2)=N_p\int_{x_2}^{x_1}dx\int_{\mathbb R}dtK_1(w_1,w_2,w)O_p(w)\,,
\ee
To evaluate the above integral, we first take $\xi_p=i\kappa_p$ with $\kappa_p>0$, so that the
time integrations are Fourier integrals, and then analytically
continue the result.
The normalization constant $N_p$ is given by
\be\label{Np}
N_p=\frac{4^{\Delta_p-1}\kappa_p}{\pi}\,.
\ee
Thus $K_i$ defined in \eqref{shadowkernel} is the integration kernel for the global OPE block. For $0<x<1$, the correlation function of the two global OPE blocks gives the global BMS
block in the normalization of \eqref{def:Fp}, i.e.
\be\label{F0-BB}
\mathcal{F}_p^{(0)}=\langle\mathcal{B}_{11p}^{\operatorname{gl}}(w_1,w_2)\mathcal{B}_{33p}^{\operatorname{gl}}(w_3,w_4)\rangle\,,
\ee
where
\be
w_1=(\infty,0)\,,\quad w_2=(1,0)\,,\quad w_3=(x,t)\,,\quad w_4=(0,0)\,.
\ee
Inserting \eqref{block-integral} into \eqref{F0-BB} gives
\be\label{shadowint}
\mathcal F^{(0)}_{ p}(x,t)=N_p^2\int_1^\infty dx_5\int_0^x dx_6\int_{\mathbb R}dt_5dt_6K_1(w_1,w_2,w_5)G_p(w_5,w_6)K_3(w_3,w_4,w_6)\,,
\ee
where $G_p$ is the two point function of operator $O_p$ given by
\be\label{internel-Gp}
G_p(w_5,w_6)=\langle O_p(w_5)O_p(w_6)\rangle=x_{56}^{-2\Delta_p}\exp\left[-2\xi_p\frac{t_{56}}{x_{56}}\right]\,.
\ee
Before proceeding to turn on boundary gravitons, we first justify the equality \eqref{shadowint} by computing 
the right hand side directly. Substituting \eqref{shadowkernel} and \eqref{internel-Gp} into \eqref{shadowint}, we can write the integrand in the following form
\be\label{integrand-global}
\Omega(x_5,x_6)\exp[i\kappa_p(tA_1+t_5A_2+t_6A_3)] \,,
\ee
where
\be
\ba
{}&\Omega(x_5,x_6)=\left[\frac{x_6(x-x_6)(x_5-1)}{x}\right]^{\Delta_p-2}x_{56}^{-2\Delta_p}\,,\\
{}&A_1=\frac{1}{x-x_6}-\frac{1}{x}\,,\quad A_2=\frac{1}{x_5-1}-\frac{2}{x_{56}}\,,\quad A_3=\frac{2}{x_{56}}+\frac{1}{x_6}+\frac{1}{x_6-x}\,.
\ea
\ee
The two time integrations can now be performed as follows
\be\label{int-t}
\int_{\mathbb R}dt_5dt_6e^{i\kappa_p(t_5A_2+t_6A_3)}=\frac{4\pi^2}{\kappa_p^2}\delta(A_2)\delta(A_3)=\frac{4\pi^2s_x^3(1-s_x)^2}{\kappa_p^2}\delta(x_5-1-s_x)\delta(x_6-1+s_x)\,.
\ee
Substituting \eqref{int-t} into \eqref{shadowint} and integrating over space coordinates gives
\be
\mathcal{F}^{(0)}_{ p}(x,t)=\frac{4^{1-\Delta_p}\pi^2N_p^2(1+s_x)^{2-\Delta_p}(1-s_x)^{\Delta_p}}{\kappa_p^2s_x}\exp\left[\frac{\xi_pt}{s_x(1-s_x^2)}\right]\,,
\ee
which agrees with \eqref{F0-iden} using \eqref{Np}.
    
\subsection{The $c_M^{-1}$ correction from boundary gravitons}
The shadow integral \eqref{shadowint} expresses the global block as two three-point functions joined by the
exchanged two-point function. In the prescription described at the beginning of this section, the $c_M^{-1}$ correction comes
from a boundary graviton propagating between these three-point functions. Let $\delta K_i$ denote the
first-order variation obtained by transforming the three primary insertions in the numerator of \eqref{shadowkernel},  while keeping its external  normalization factor fixed. The quantity to evaluate is
\be\label{F1-KK}
\delta\mathcal{F}_p=N_p^2\int_1^\infty dx_5\int_0^x dx_6\int_{\mathbb R}dt_5dt_6G_p(w_5,w_6)\langle \delta K_1(w_1,w_2,w_5)\delta K_3(w_3,w_4,w_6)\rangle\,.
\ee
The expectation value uses the quadratic boundary graviton propagator on the BMS vacuum orbit derived in \cite{Merbis:2019wgk}. The pairing $G_p$, integration measure, and integration domains are not varied. Since the propagator is of order $c_M^{-1}$, so will be $\delta\mathcal F_p$. After evaluating
\eqref{F1-KK}, we will compare it with the correction \eqref{F1:pairiden} derived earlier.

To compute $\delta K_i$, 
we perform a BMS transformation \eqref{bms-transformation}. Plugging \eqref{def:primary} into \eqref{shadowkernel} gives the transformed kernel
\be\label{Kf}
K_i^{(f,h)}(w_1,w_2,w_3)
={}\frac{x_{12}^{2\Delta_i}e^{2\xi_i t_{12}/x_{12}}}
{c_{i\tilde p i}}
j_i(w_1)j_i(w_2)j_{\tilde p}(w_3)\langle O_i(\widetilde w_1)
O_{\tilde p}(\widetilde w_3)O_i(\widetilde w_2)\rangle\,,
\ee
where
\be
j_i(w)=[f'(x)]^{\Delta_i}\exp\left[\xi_i\frac{f''(x)t+h'(x)}{f'(x)}\right]\,.
\ee
Writing the infinitesimal transformation as
\be
f(x)=x+\epsilon(x)\,,\quad h(x)=\alpha(x)\,,
\ee
we can expand \eqref{Kf} up to linearized order
\be
K_i^{(f,h)}(w_1,w_2,w_3)= K_i(w_1,w_2,w_3)+\delta K_i(w_1,w_2,w_3)+\cdots\,,
\ee
where we have omitted terms which are at least quadratic in the fluctuations.
The linearized part $\delta K_i$ is computed to be
\be\ba\label{deltaK}
\frac{\delta K_i(w_1,w_2,w_3)}{K_i(w_1,w_2,w_3)}=&\Delta_iJ_{12}+\xi_iH_{12}+\frac{2-\Delta_p}{2}(J_{13}+J_{23}-J_{12})-\frac{\xi_p}{2}(H_{13}+H_{23}-H_{12})\,,
\ea
\ee
where
\be\ba\label{def:JH}
J_{ij}=&\epsilon'(x_i)+\epsilon'(x_j)-\frac{2[\epsilon(x_i)-\epsilon(x_j)]}{x_i-x_j}\,,\\
H_{ij}=&\alpha'(x_i)+\alpha'(x_j)+t_i\epsilon''(x_i)+t_j\epsilon''(x_j)
-\frac{2[\alpha(x_i)-\alpha(x_j)+t_i\epsilon'(x_i)-t_j\epsilon'(x_j)]}{x_{i}-x_j}\\
&+\frac{2(t_i-t_j)[\epsilon(x_i)-\epsilon(x_j)]}{(x_i-x_j)^2}
\ea
\ee
We evaluate the correlator in \eqref{F1-KK} directly in position space. With the global modes removed, the mixed vacuum propagator on the plane is given by \footnote{Correlator \eqref{cor-ea}  follows from the cylinder propagator (5.42b) in \cite{Merbis:2019wgk}. Under the BMS transformation $x=e^{-i\varphi},t=-iue^{-i\varphi}$, the vector components at cylinder time zero transform as  $\epsilon=-ix\epsilon_{\operatorname{cyl}}(\varphi)$ and $\alpha(x)=-ix\tilde\alpha_{\operatorname{cyl}}(\varphi)$. The two factors of $-ix$ give the prefactor $-x_1x_2$.}
\be\label{cor-ea}
g(x_1,x_2)\equiv\langle\epsilon(x_1)\alpha(x_2)\rangle=\frac{-3x_1x_2}{c_M}\left[3\zeta-2-\frac{2(1-\zeta)^2}{\zeta}\log(1-\zeta)\right]\,,
\ee
where $\zeta=\frac{x_2}{x_1}<1$ and the other two correlators vanish at order $c_M^{-1}$.  Combining \eqref{deltaK}, \eqref{def:JH} and \eqref{cor-ea}, it can be shown   that
\be\label{KKcorrelator}
\frac{\langle\delta K_1\delta K_3\rangle}{K_1K_3}= 
\frac{12}{c_M}\left(\Delta_1\partial_{\xi_1}+\Delta_3\partial_{\xi_3}+(\Delta_p-2)\partial_{\xi_p}+t_5\partial_{x_5}+t\partial_x+t_6\partial_{x_6}\right)\mathcal{C}\,,
\ee
where 
\be
\ba
\mathcal C={}&
\xi_1\xi_3
\left(
-2+\frac{x-2}{x}\log(1-x)
\right)
\\
&+\xi_1\xi_p
\left(
1+
\frac{x_6(1-x)}{x(x-x_6)}\log(1-x)
+
\frac{x_6^2-2x_6+x}{x_6(x-x_6)}\log(1-x_6)
\right)
\\
&+\xi_3\xi_p
\left(
1+
\frac{1-x}{x(x_5-1)}\log(1-x)
+
\frac{x_5^2-2x_5+x}{x(x_5-1)}
\log\left(1-\frac{x}{x_5}\right)
\right)
\\
&+\xi_p^2
\left[
-\frac12
-\frac{x_6(1-x)^2}{2x(x-x_6)(x_5-1)}\log(1-x)
\right.
\\
&\quad
+\frac{(x_6-1)(x_6x-2x_6+x)}
 {2x_6(x-x_6)(x_5-1)}\log(1-x_6)
\\
&\quad
+\frac{x_6(x-x_5)(x_5+x-2)}
 {2x(x-x_6)(x_5-1)}
 \log\left(1-\frac{x}{x_5}\right)
\\
&\quad\left.
-\frac{x_6^2x-2x_6^2-2x_6x_5^2+4x_6x_5+x_5^2x-2x_5x}
 {2x_6(x-x_6)(x_5-1)}
 \log\left(1-\frac{x_6}{x_5}\right)
\right].
\ea\ee
Having determined the expectation value in \eqref{KKcorrelator}, the remaining integrations follow the same localization as the global block, except that the integrand now contains terms linear in $t_5$ and $t_6$, which lead to derivatives of delta functions after integration. 
Combining \eqref{integrand-global} with \eqref{KKcorrelator}, the integrand in \eqref{F1-KK} can be formally written as
\be
\Omega(x_5,x_6)\exp[i\kappa_p(tA_1+t_5A_2+t_6A_3)](C_0+t_5C_1+t_6C_2)
\ee
with coefficients $C_0,C_1,C_2$ being independent of $t_5,t_6$. 
More explicitly, they are given by
\be
\ba
C_0&=\frac{12}{c_M}(\Delta_1\p_{\xi_1}+\Delta_3\p_{\xi_3}+(\Delta_p-2)\p_{\xi_p}+t\p_x)\mathcal{C}\,,\quad
C_1=\frac{12\p_{x_5}\mathcal{C}}{c_M}\,,\quad
C_2=\frac{12\p_{x_6}\mathcal{C}}{c_M}\,.
\ea
\ee
After integrating over time coordinates, we get
\be\ba
{}&\int_{\mathbb R}dt_5dt_6\exp[i\kappa_p(t_5A_2+t_6A_3)](C_0+t_5C_1+t_6C_2)\\
=&\frac{4\pi^2}{\kappa_p^2}\left[C_0\delta(A_2)\delta(A_3)+\frac{C_1}{i\kappa_p}\delta'(A_2)\delta(A_3)+\frac{C_2}{i\kappa_p}\delta(A_2)\delta'(A_3)\right]\,.
\ea
\ee
A useful simplification that occurs at the localization point $(x_5)_*=1+s_x,(x_6)_*=1-s_x$ is
\be\label{local-eq}
\mathcal{C}|_{*}=Q(x)\,,\quad \p_{x_5}\mathcal{C}\big|_{*}=\p_{x_6}\mathcal{C}\big|_{*}=0\,,
\ee
where $Q(x)$ is the function in \eqref{functions:PQ}.  
Consequently, integrating over $x_5$ and $x_6$ and using integration by parts leads to
\be
\ba\label{deltaF-integration}
\frac{\delta\mathcal{F}_p}{\mathcal{F}_p^{(0)}}=\frac{12}{c_M}\left\{[\Delta_1\p_{\xi_1}+\Delta_3\p_{\xi_3}+(\Delta_p-2)\p_{\xi_p}]Q+t\p_x\mathcal{C}|_{*}-\frac{1}{\xi_p}[\p_{A_2}\p_{x_5}+\p_{A_3}\p_{x_6}]\mathcal{C}|_{*}\right\}\,.
\ea
\ee
Using the relation
\be
\left.\partial_{A_2}\right|_*
=\frac{s_x}{2}\left[-(1+s_x)^2\partial_{x_5}
+(1-s_x)^2\partial_{x_6}\right]\,,\quad \left.\partial_{A_3}\right|_*
=\frac{s_x(1-s_x)^2}{2}
\left(\partial_{x_5}-\partial_{x_6}\right)\,,
\ee
we find
\be\label{eq1}
\frac{1}{\xi_p}
\left(\partial_{A_2}\partial_{x_5}
+\partial_{A_3}\partial_{x_6}\right)\mathcal C\big|_*
=\left(\frac12\partial_{\xi_1}
+\frac12\partial_{\xi_3}-\partial_{\xi_p}\right)Q\,.
\ee
Moreover, it is easy to see that 
\be\label{eq2}
\p_x\mathcal C|_*=Q'(x)\ee
by virtue of \eqref{local-eq}. Plugging \eqref{eq1} and \eqref{eq2} into \eqref{deltaF-integration}, we find
\be
\delta\mathcal F_p=\frac{12}{c_M}\mathcal{F}_p^{(0)}\left[P(x)+tQ'(x)\right]
\ee
with functions $P$ and $Q$ given by \eqref{functions:PQ}. The integral therefore reproduces \eqref{F1:pairiden}, giving $\delta\mathcal{F}_p=c_M^{-1}\mathcal{F}_p^{(1)}$ for two
pairs of identical external operators.

\section{Comparison with the ultra-relativistic limit of Virasoro blocks}\label{sec:ur}
This section gives a check of the result for two identical external pairs by taking the UR limit of 
the known Virasoro block at order $c^{-1}$. The construction uses a tensor
product of two chiral Virasoro modules, one of which is flipped. We first expand the Virasoro block at large central charge to collect the leading and subleading terms. At each order, we then take the UR limit and show the results agree with the BMS block at the corresponding order in $c_M$.
We then prove that, at each fixed total descendant level, the formal UR
contraction commutes with extraction of the $c_M^{-1}$ coefficient.  
 
\subsection{Flipped right-moving module and contraction dictionary}
Let $\mathcal{L}_n$ and $\bar{\mathcal{L}}_n$ generate the two Virasoro algebras with central charges $c,\bar c$ of the parent CFT. The UR contraction is obtained by taking the limit $\epsilon\to0$ of the following relations \cite{Bagchi:2012cy}
\be\label{UR}
L_n=\mathcal{L}_n-\bar{\mathcal{L}}_{-n}\,,\quad M_n=\epsilon(\mathcal{L}_n+\bar{\mathcal{L}}_{-n})\,.
\ee
Under this limit, $(L_n,M_n)$ satisfy the BMS$_3$ algebra with central charges given by
\be\label{UR:central}
c_L=c-\bar c\,,\quad c_M=\epsilon(c+\bar c)\,.
\ee
From \eqref{UR}, it is easy to see that a conventional highest weight module in both left and right moving sectors does not contract directly to the highest-weight BMS$_3$ module. Introduce instead the flipped right-moving generators $\tilde{\mathcal L}_n=-\bar{\mathcal L}_{-n}$. This automorphism and its role in contractions to highest-weight
BMS representations are discussed in \cite{Bagchi:2019unf,Hao:2021urq}. 
Under this automorphism, $\tilde{\mathcal L}_n$ generates  a Virasoro algebra with central charge $\tilde{\bar c}=-\bar c$. In terms of the flipped generators, \eqref{UR} and \eqref{UR:central} become
\be\ba\label{rel:BMS-flipCFT}
L_n=&\mathcal{L}_n+\tilde{\mathcal L}_n\,,\quad M_n=\epsilon(\mathcal{L}_n-\tilde{\mathcal L}_n)\,,\\
c_L=&c+\tilde{\bar c}\,,\quad c_M=\epsilon(c-\tilde{\bar c})\,.
\ea
\ee
where $\epsilon\to0$ is left implicit. Hence, the highest weight representation of BMS$_3$ algebra can be obtained from the UR limit of the highest weight representation of two copies of Virasoro algebra formed by $(\mathcal{L}_n,\tilde{\mathcal L}_n)$.

\subsection{Virasoro block at order $1/c$} 
Consider the  chiral Virasoro block for two  identical external pairs of primaries with weights $h_1,h_3$ and an exchanged primary of weight $h_p$.  The contribution of this exchanged module can be written as 
\be\label{def:CFTblock}
\langle O_1(\infty)O_1(1)\Pi_{h_p}^{\operatorname{Vir}}O_3(z)O_3(0)\rangle=C_{11h_p}C_{33h_p}z^{h_p-2h_3}G_{h_p}(c;z)\,,
\ee
where $\Pi_{h_p}^{\operatorname{Vir}}$ is the projector onto the exchanged primary and all its Virasoro descendants. 
At fixed weights, $G_{h_p}$ has the following large $c$ expansion \cite{Bombini:2018jrg}
 \be\label{CFT-largec}
 G_{h_p} (c;z)=F_{h_p}(z)+\frac{12}{c}[h_1h_3f_a(z)+(h_1+h_3)f_b(z)+f_c(z)]+O(c^{-2})\,,
 \ee   
where $F_{h_p}(z)={}_2F_1(h_p,h_p,2h_p;z)$ and the three functions in the subleading terms are
\be\ba\label{functions:f}
f_a(z)&=-\left[2+\frac{(2-z)\log(1-z)}{z}\right] F_{h_p}(z)\,,\\
f_b(z)&=h_pF_{h_p}(z)+\frac{h_p(1-z)\log(1-z)}{z}{}_2F_1(h_p,h_p+1,2h_p;z)\,,\\
f_c(z)&=-\frac{h_p^2}{2}F_{h_p}(z)-\frac{h_p^2(1-z)^2\log(1-z)}{2z}{}_2F_1(h_p+1,h_p+1,2h_p;z)+\frac{h_p(h_p-1)}{2}\p_{h_p}F_{h_p}(z)\,.
\ea\ee
A conformal block of a CFT$_2$ is the product of left and right moving Virasoro blocks.  We use $\iota=1$ to label quantities in the left moving sector and $\iota=-1$ to label   the right moving sector. For each $\alpha\in\{1,3,p\}$, by virtue of \eqref{rel:BMS-flipCFT}, we define
\be\label{CFTparameters}    
h_{\alpha,\iota}=\frac{1}{2}\left(\Delta_\alpha+\iota\frac{\xi_\alpha}{\epsilon}\right)\,,\quad c_\iota=\frac{1}{2}\left(c_L+\iota\frac{c_M}{\epsilon}\right)\,,\quad z_\iota=x+\iota\epsilon t\,.
\ee
Here $z_\iota$ denotes the cross ratio of the four chiral insertion points $(\infty,1,x+\iota\epsilon t,0)$, with $\iota=-1$ referring to the flipped sector.
Although the chiral weights diverge as $\epsilon^{-1}$, their ratios to the corresponding central charges 
\be
\frac{h_{\alpha,\iota}}{c_\iota}=\frac{\xi_\alpha}{c_M}+O(\epsilon)
\ee
remain small for large $c_M$. For every fixed nonzero $\epsilon$, the weights are fixed as
$c_M\to\infty$, so the expansion \eqref{CFT-largec} applies. The interchange
with the contraction is justified below at each fixed level.

To match the normalization of the BMS block in
\eqref{def:Fp}, we consider the following quantity
\be\label{CFTblock}
\mathcal F_\epsilon(x,t)
=\prod_{\iota=\pm1}
z_\iota^{h_{p,\iota}}
G_{h_{p,\iota}}(c_\iota;z_\iota)\,.
\ee
To isolate the relative correction, we write
\be\label{Sigma}
\Sigma(h_p,h_1,h_3;z)=\frac{h_1h_3f_a(z)+(h_1+h_3)f_b(z)+f_c(z)}{F_{h_p}(z)}\,.
\ee
Expanding \eqref{CFTblock} at large $c_M$ for fixed nonzero $\epsilon$ gives
\be\ba\label{CFTblock-largec}
\mathcal{F}_\epsilon(x,t)&=\mathcal{F}^{(0)}_\epsilon(x,t)\left[1+\frac{24\epsilon}{c_M}\sum_{\iota=\pm1}\iota\Sigma(h_{p,\iota},h_{1,\iota},h_{3,\iota};z_\iota)+O(c_M^{-2})\right]\,,\\
\mathcal{F}^{(0)}_\epsilon(x,t)&=\prod_{\iota=\pm1}z_\iota^{h_{p,\iota}}F_{h_{p,\iota}}(z_\iota)\,.
\ea
\ee
In the following, we will show that, as $\epsilon\to0$,     $\mathcal{F}^{(0)}_\epsilon(x,t)$ and the subleading term in the bracket have  well defined limits, which agree with $\mathcal{F}^{(0)}_{ p}$ and $\frac{\mathcal{F}^{(1)}_{ p}}{c_M\mathcal{F}^{(0)}_{p}}$  respectively.
\subsection{UR limit and comparison}
By virtue of \eqref{CFTparameters}, we first need to find the asymptotic behaviour of $F_{h}(z)$ at $|h|\to\infty$. Set
\be
s_z=\sqrt{1-z}\,,\quad \rho_z=\frac{1-s_z}{1+s_z}\,.
\ee
Two standard transformations of the hypergeometric function give
\be\label{eq:hyper}
F_h(z)=(1+\rho_z)^{2h}{}_2F_1\left(\frac{1}{2},h,h+\frac{1}{2};\rho^2_z\right)=\frac{(1+\rho_z)^{2h}}{\sqrt{1-\rho_z^2}}{}_2F_1\left(\frac{1}{2},\frac{1}{2},h+\frac{1}{2};\frac{\rho_z^2}{\rho_z^2-1}\right)\,.
\ee
On the OPE branch, we have $\text{Re}~s_z>0\Leftrightarrow|\rho_z|<1$. As a result, it can be checked that 
\be
\text{Re}\left(\frac{\rho_z^2}{\rho_z^2-1}\right)<\frac{1}{2}\,.
\ee
In this region, the large $|h|$ behavior of the hypergeometric function on the right hand side of \eqref{eq:hyper} has the following asymptotic expansion
\be\label{hyper-largeh}
{}_2F_1\left(\frac{1}{2},\frac{1}{2},h+\frac{1}{2};\frac{\rho_z^2}{\rho_z^2-1}\right)=1-\frac{\rho^2_z}{4h(1-\rho_z^2)}+O(h^{-2})\,,
\ee
which holds as long as $|h+\frac{1}{2}+n|\geq\delta$ for a fixed sufficiently small positive constant $\delta$ and $n\in\mathbb N$. We assume this condition is met as $h\to-\infty$. Plugging \eqref{hyper-largeh} into \eqref{eq:hyper} gives
\be\label{Fh-largeh}
F_h(z)=\frac{(1+\rho_z)^{2h}}{\sqrt{1-\rho_z^2}}\left[1-\frac{\rho^2_z}{4h(1-\rho_z^2)}+O(h^{-2})\right]\,.
\ee
Substituting \eqref{Fh-largeh} and \eqref{CFTparameters} into the second equation in \eqref{CFTblock-largec} and taking the limit $\epsilon\to0$, we get
\be
\lim_{\epsilon\to0}\mathcal{F}^{(0)}_\epsilon(x,t)=4^{\Delta_p-1}\frac{(1+s_x)^{2-\Delta_p}(1-s_x)^{\Delta_p}}{s_x}e^{\frac{\xi_pt}{s_x(1-s_x^2)}}=\mathcal{F}^{(0)}_{ p}(x,t)\,.
\ee
Therefore, the UR limit of the leading CFT block agrees with the leading BMS block.
To compute the UR limit of the first subleading order, we first rewrite $\Sigma$ in terms of $F_h(z)$ using the following identities
\be\label{iden-Fh}
{}_2F_1(h,h+1,2h;z)=F_h(z)+\frac{z}{h}\p_zF_h(z)\,,\quad {}_2F_1(h+1,h+1,2h;z)=\frac{(h+z\p_z)^2}{h^2}F_h(z)\,.
\ee
Substituting \eqref{Fh-largeh} and \eqref{iden-Fh} into \eqref{functions:f} gives
\be
\ba\label{functions:f-largeh}
\frac{f_a(z)}{F_{h_p}(z)}&=-2-\frac{2(1+s_z^2)}{1-s_z^2}\log s_z\,,\\
\frac{f_b(z)}{F_{h_p}(z)}&=\left(1+\frac{2s_z}{1-s_z^2}\log s_z\right)h_p+\frac{1-s_z}{2(1+s_z)}\log s_z+O(h_p^{-1})\,,\\
\frac{f_c(z)}{F_{h_p}(z)}&=\left(\log\frac{2}{1+s_z}-\frac{1}{2}-\frac{s_z^2}{1-s_z^2}\log s_z\right)h^2_p-\left(\log\frac{2}{1+s_z}+\frac{s_z}{1+s_z}\log s_z\right)h_p+O(1)\,.
\ea
\ee
Plugging \eqref{functions:f-largeh} into \eqref{Sigma} and using \eqref{CFTparameters}, it can be checked directly that
\be
\Sigma(h_{p,\iota},h_{1,\iota},h_{3,\iota};z_\iota)=\frac{Q(z_\iota)}{4\epsilon^2}+\frac{\iota P(z_\iota)}{4\epsilon}+O(1)\,.
\ee
As a consequence,
\be
\lim_{\epsilon\to0}\frac{24\epsilon}{c_M}\sum_{\iota=\pm1}\iota\Sigma(h_{p,\iota},h_{1,\iota},h_{3,\iota};z_\iota)=\frac{12}{c_M}[P(x)+tQ'(x)]=\frac{\mathcal{F}^{(1)}_{ p}}{c_M\mathcal{F}^{(0)}_{ p}}\,.
\ee
To conclude, we study the UR limit $\epsilon\to0$ of the conformal block through the first two orders in the large $c$ expansion. The result agrees precisely with the BMS block at the first two orders in the large $c_M$ expansion.

\paragraph{Comments on the order of limits}
Given the exact conformal block \eqref{CFTblock}, we should in principle take the limit $\epsilon\to0$ first and then perform the large $c_M$ expansion to compare with the intrinsic calculations. In the previous discussion, however, we first expand $\mathcal{F}_\epsilon$ in large $c_M$ and then take the UR limit at each order. Such agreement is not guaranteed unless the two operations commute. We now prove this commutativity at
every fixed descendant level. Consider the linear combination of \eqref{UR} and denote $L_n^{(\epsilon)},M_n^{(\epsilon)}$ as the corresponding generators at nonzero $\epsilon$.
The algebra formed by $L_n^{(\epsilon)},M_n^{(\epsilon)}$ is given by
\be
\ba\label{BMS-epsilon}
[L_n^{(\epsilon)},L_m^{(\epsilon)}]&=(n-m)L_{n+m}^{(\epsilon)}+\frac{c_L}{12}n(n^2-1)\delta_{n+m,0}\,,\\
[L_n^{(\epsilon)},M_m^{(\epsilon)}]&=(n-m)M_{n+m}^{(\epsilon)}+\frac{c_M}{12}n(n^2-1)\delta_{n+m,0}\,,\\
[M_n^{(\epsilon)},M_m^{(\epsilon)}]&=\epsilon^2\left((n-m)L_{n+m}^{(\epsilon)}+\frac{c_L}{12}n(n^2-1)\delta_{n+m,0}\right)\,.
\ea
\ee
where $c_L,c_M$ are defined through \eqref{UR:central}. At $\epsilon=0$, \eqref{BMS-epsilon} is exactly the BMS$_3$ algebra.
For $\epsilon\ne0$, it is isomorphic to the direct sum of two Virasoro algebras.
Fix the ordered PBW words in $L_{-n}$ and $M_{-n}$ at the total level $N$. For $\epsilon\neq0$, these words acting on the primary state form a basis of the level $N$ subspace of the tensor-product Virasoro Verma module
\be
\bigoplus_{N_++N_-=N}\mathcal{V}^{(+)}_{p,N_+}\otimes\mathcal V^{(-)}_{p,N_-}\,.
\ee
Let $ G_N(\epsilon,c_M)$ be the exact Gram matrix in this basis and $\mathbf l_N(\epsilon),\mathbf r_N(\epsilon)$ be the corresponding normalized three-point vectors. The contribution from the level-$N$ sector to the CFT block is then given by
\be
\mathcal{F}_{\epsilon,N}=\mathbf l_N(\epsilon)^TG_N(\epsilon,c_M)^{-1}\mathbf r_N(\epsilon)\,.
\ee
The three-point  vectors are independent of the central charge $c_M$. Moreover, it can be checked that both the primary three-point functions and the Ward operators associated with $L_n^{(\epsilon)}$ and $M_n^{(\epsilon)}$ are holomorphic near $\epsilon=0$. At any fixed descendant level, the three-point vector components follow from finitely many applications of these operators. As a result, both normalized three-point vectors are holomorphic in $\epsilon$. 

Entries of the Gram matrix are polynomials in $\epsilon$ and $c_M$, as follows
directly from the deformed commutation relations \eqref{BMS-epsilon}. 
Order the PBW basis by degree,  defined as the number of generators $L_{n\leq-2}^{(\epsilon)},M_{n\leq-2}^{(\epsilon)}$,  as in section \ref{sec:degree}. Since $c_M$ only arises from the commutator between $L_n$ and $M_{-n}$ with $|n|\geq2$, the same counting of
central contractions gives
$\deg_{c_M} A_N^{ij}(\epsilon,c_M)\leq\min(i,j)$, where $A_N^{ij}(\epsilon,c_M)$ is the block formed by inner products between degree $i$ and $j$ subspaces. 
Rescaling the rows of degree $i$ therefore gives the matrix
\begin{equation}
 \widehat G_N(\epsilon,c_M^{-1})
 =\left(c_M^{-i}A_N^{ij}(\epsilon,c_M)\right)_
 {0\leq i,j\leq\lfloor N/2\rfloor}\,,
\end{equation}
whose entries are polynomials in $\epsilon$ and $c_M^{-1}$.
At $(\epsilon,c_M^{-1})=(0,0)$, the blocks below the diagonal vanish,
and the diagonal blocks are $\Gamma_{N,i}$. Consequently,
\begin{equation}
 \det\widehat G_N(0,0)
 =\prod_{i=0}^{\lfloor N/2\rfloor}\det\Gamma_{N,i}\ne0\,,
\end{equation}
as we have shown in section \ref{sec:degree}.
The inverse of $\widehat G_N$ is therefore holomorphic in a neighborhood
of the origin. The exact inverse Gram matrix is
\begin{equation}
 G_N(\epsilon,c_M)^{-1}
 =\widehat G_N(\epsilon,c_M^{-1})^{-1}
 \operatorname{diag}_{0\leq i\leq\lfloor N/2\rfloor}
 \left(c_M^{-i}\mathbf 1_{d_{N,i}}\right),
\end{equation}
where $d_{N,i}$ is the dimension defined in \eqref{def:dimension}. Hence this inverse,
and therefore $\mathcal F_{\epsilon,N}$, are jointly holomorphic in
$(\epsilon,c_M^{-1})$ near $(0,0)$.

This proves the required commutativity
at every fixed descendant level.
Evaluation at $\epsilon=0$ consequently commutes with taking the
Taylor coefficient of $c_M^{-1}$ and  setting $\epsilon=0$ at each order of the large $c_M$ expansion gives the intrinsic BMS contribution at level $N$.

\appendix
\section{Normalized global-descendant three-point vectors.}\label{app:3pv}
In this section, we provide explicit expressions for $\mathbf l_q^{\text{gl}}$ and $\mathbf r_q^{\text{gl}}$ by deriving the differential relations \eqref{rel:vec-W} and evaluating their components.
For the left  vector, translating the exchanged operator from the origin to an arbitrary point gives
\be
O_p(x,t)=e^{xL_{-1}+tM_{-1}}O_p(0,0)e^{-xL_{-1}-tM_{-1}}\,.
\ee
Consequently, the function $\mathcal{W}_{\Delta_p,\xi_p}$ defined by \eqref{def:W} can be written as
\be
\mathcal{W}_{\Delta_p,\xi_p}(x,t)=\sum_{a,b\geq0}\frac{x^at^b}{a!b!}\frac{\langle O_1(\infty,0)O_2(1,0)L^a_{-1}M^b_{-1}| p\rangle}{c_{12p} }\,.
\ee
Setting $b=q-a$ and using \eqref{Commutator} gives the components of $\mathbf l_q^{\text{gl}}$
\be
(\mathbf l_q^{\text{gl}})_a(\Delta_p,\xi_p)=\p^a_x\p^{q-a}_t\mathcal{W}_{\Delta_p,\xi_p}(x,t)\Big|_{x=t=0}=(\Delta_p-\Delta_{12}+q-a)_a(\xi_p-\xi_{12})^{q-a}\,,
\ee
where $(x)_n=\frac{\Gamma(x+n)}{\Gamma(x)}$ is the Pochhammer symbol.
For the right vector, the $a$th component contains $((D_q)_a)^\dagger=M_1^{q-a}L_1^a$. Commuting these positive
modes through $O_3(x,t)O_4(0,0)$ gives the Ward operators in \eqref{Commutator}. Note that the function $\widetilde {\mathcal W}$ defined in \eqref{def:W} contains an external factor. Conjugating the differential operators in the Ward identities by this factor gives
\be\ba\label{right-MLW}
(\mathbf r_q^{\text{gl}})_a(\Delta_p,\xi_p)&=\widehat M_1^{q-a}\widehat L_1^a \widetilde {\mathcal W}_{\Delta_p,\xi_p}\,,
\ea
\ee
where
\be\label{ML-right}
\widehat M_1=x^2\p_t+\xi_{34}x\,,\quad \widehat L_1=x^2\p_x+2xt\p_t+\Delta_{34}x+\xi_{34}t\,.
\ee
Plugging \eqref{ML-right} into \eqref{right-MLW} and using \eqref{def:W}, the right global vector is computed to be
\be\ba
(\mathbf r_q^{\text{gl}})_a(\Delta_p,\xi_p)=a!x^{\Delta_p+q}e^{\frac{\xi_pt}{x}}(\xi_p+\xi_{34})^{q-a}L_a^{(\Delta_p+\Delta_{34}+q-a-1)}\left(-\frac{(\xi_p+\xi_{34})t}{x}\right) \,, 
\ea
\ee
where $L_a^{(\alpha)}(z)$ is the generalized Laguerre polynomial.

\section{Construction of $|U_m\rangle$ and $|V_m\rangle$}
In this section, we solve the conditions in \eqref{condition}.
Consider the following ansatz for $|U_m\rangle$
\be
|U_m\rangle=f_m(\{M_{-n}\};\xi_p)| p\rangle\,,
\ee
where $f_m(\{M_{-n}\};\xi_p)$ is an $M$-polynomial given by
\be
f_m(M;\xi_p)=\sum_{k=1}^ma_{m,k}M_{-k}M_{-1}^{m-k}\,,\quad a_{m,m}=1.
\ee
The conditions $M_1|U_m\rangle=0,M_0|U_m\rangle=\xi_p|U_m\rangle$ are automatic since all $M$ modes commute. The action of $L_1$ on a pure $M$-polynomial is given by the operator
\be
L_1f_m|p\rangle=\left(2\xi_p\frac{\p}{\p M_{-1}}+\sum_{n\geq2}(n+1)M_{-n+1}\frac{\p}{\p M_{-n}}\right)f_m|p\rangle\,.
\ee
Therefore, $L_1|U_m\rangle=0$ implies
\be
(k+2)a_{m,k+1}+2\xi_p(m-k+\delta_{k,1})a_{m,k}=0\,,\quad 1\leq k\leq m-1\,.
\ee
The recurrence has the solution
\be\label{coe:a}
a_{m,k}=\left\{\begin{array}{ll}
  (-1)^{m-k}\frac{(m+1)!}{(k+1)!(m-k)!(2\xi_p)^{m-k}}   &\quad 2\leq k\leq m  \\
   (-1)^{m-1}\frac{m^2-1}{2(2\xi_p)^{m-1}}  &\quad k=1
\end{array}\right. \,.
\ee
To solve for $|V_m\rangle$, we first consider the ansatz
\be\label{Vm0}
|V_m^{(0)}\rangle=\sum_{n\geq1}L_{-n}\frac{\p f_m}{\p M_{-n}}|p\rangle\,.
\ee
 This state already satisfies the two conditions involving $M_1$ and $M_0$. More explicitly, we have
\be
M_1|V_m^{(0)}\rangle=\sum_{n\geq1}(n+1)M_{-n+1}\frac{\p f_m}{\p M_{-n}}|p\rangle=L_1|U_m\rangle=0\,,
\ee
 and
\be\ba
(M_0-\xi_p)|V_m^{(0)}\rangle=&\sum_{n=1}^m[M_0,L_{-n}]\frac{\p f_m}{\p M_{-n}}|p\rangle+\sum_{n=1}^mL_{-n}(M_0-\xi_p)\frac{\p f_m}{\p M_{-n}}|p\rangle\\
=&\sum_{n=1}^mnM_{-n}\frac{\p f_m}{\p M_{-n}}|p\rangle=mf_m|p\rangle=m|U_m\rangle\,.
\ea\ee
However, $|V_m^{(0)}\rangle$ is not annihilated by $L_1$. Actually,  differentiating the explicit polynomial identity $L_1f_m|p\rangle=0$
with respect to $M_{-n}$ gives
\be\ba
0=&\left[2\xi_p\frac{\p}{\p M_{-1}}\frac{\p f_m}{\p M_{-n}}+\sum_{k=2}^m(k+1)M_{-k+1}\frac{\p}{\p M_{-k}}\frac{\p f_m}{\p M_{-n}}+(n+2)\frac{\p f_m}{\p M_{-n-1}}\right]|p\rangle\\
=&\left[L_1\frac{\p f_m}{\p M_{-n}}+(n+2)\frac{\p f_m}{\p M_{-n-1}}\right]|p\rangle\,.
\ea
\ee
Therefore,
\be\ba
L_1|V_m^{(0)}\rangle=&\sum_{n=1}^m[L_1,L_{-n}]\frac{\p f_m}{\p M_{-n}}|p\rangle+\sum_{n=1}^mL_{-n}L_1\frac{\p f_m}{\p M_{-n}}|p\rangle\\
=&2L_0\frac{\p f_m}{\p M_{-1}}|p\rangle+\sum_{n=2}^m(n+1)L_{1-n}\frac{\p f_m}{\p M_{-n}}|p\rangle-\sum_{n=1}^{m-1}(n+2)L_{-n}\frac{\p f_m}{\p M_{-n-1}}|p\rangle\\
=&2L_0\frac{\p f_m}{\p M_{-1}}|p\rangle=2(\Delta_p+m-1)\frac{\p f_m}{\p M_{-1}}|p\rangle\,.
\ea
\ee
To construct $|V_m\rangle$, we notice that by differentiating $L_1f_m|p\rangle=0$ with respect to $\xi_p$, we get
\be
L_1\frac{\p f_m}{\p\xi_p}|p\rangle=-2\frac{\p f_m}{\p M_{-1}}|p\rangle\,.
\ee
Consequently, the state
\be\label{Vm}
|V_m\rangle=|V_m^{(0)}\rangle+(\Delta_p+m-1)\p_{\xi_p}f_m|p\rangle
\ee
is obviously annihilated by $L_1$. Moreover, the second term is annihilated by $M_1$ and $M_0-\xi_p$. Therefore, \eqref{Vm} satisfies all the requirements in \eqref{condition}.
\section{BMS block at $\xi_p=0$}\label{sec:xip=0}
The representation theory at $\xi_p=0$ is discontinuous from that at nonzero $\xi_p$ \cite{Chen:2022jhx}. In this appendix, we first
determine the global block after quotienting by the null submodule, then prove that the truncation
at large $c_M$ survives the quotient, and finally give the linearized correction.  At level one, the Gram matrix reads
\be
\mathcal{G}_1^p=\left(\begin{array}{cc}
   2\Delta_p  & 2\xi_p \\
   2\xi_p  & 0
\end{array}\right)\,.
\ee
Thus, the state $M_{-1}|p\rangle$ is null when $\xi_p=0$. The coefficients in \eqref{coe:a} also diverge in this limit, so neither the inverse Gram matrix nor the basis built from $|U_m\rangle$ and $|V_m\rangle$ may be obtained by setting $\xi_p=0$ in section \ref{sec:intrinsic}. Instead, we need to take the non-vacuum singlet quotient by the null submodule generated by $M_{-1}|p\rangle$\,. 

Inserting $M_{-1}|p\rangle$  into the two OPEs and applying the Ward identity gives
\be
\xi_1=\xi_2\,,\quad \xi_3=\xi_4\,.
\ee
After quotienting the $M_{-1}$ tower, the global module is
the ordinary $SL(2,\mathbb R)$ highest-weight module. In the normalization of \eqref{OPE}, the global block is given by
\be
\mathcal{F}^{(0)}_{\Delta_p}(x,t)=x^{\Delta_p}{}_2F_1(\Delta_{p2,1},\Delta_{p3,4},2\Delta_p,x)\,.
\ee
To extract the coefficient of $c_M^{-1}$ at level $N$, we  choose  the global descendant state and the descendant states containing one non-global generator. They are given by
\be
|C_N\rangle=L_{-1}^N|p\rangle\,,\quad |A_{N,k}\rangle=L_{-k}L_{-1}^{N-k}|p\rangle\,,\quad |B_{N,k}\rangle=L_{-1}^{N-k}M_{-k}|p\rangle\,,\quad k=2,\cdots,N\,.
\ee
The norm of global states and their overlaps with degree one states are given by
\be
\langle C_N|C_N\rangle=N!(2\Delta_p)_N\,,\quad \langle C_N|A_{N,k}\rangle=N!(2\Delta_p)_{N-k}[(k+1)\Delta_p+N-k]\,,\quad \langle C_N|B_{N,k}\rangle=0\,.
\ee
Define
\be\label{hatA}
|\hat A_{N,k}\rangle=|A_{N,k}\rangle-\alpha_{N,k}|C_N\rangle\,,\quad \alpha_{N,k}=\frac{(k+1)\Delta_p+N-k}{(2\Delta_p+N-k)_k}\,,
\ee
then we have $\langle C_N|\hat A_{N,k}\rangle=0$. In the large $c_M$ limit, the leading mixed Gram block matrix is lower triangular with entries given by
\be
\langle\hat A_{N,k}|B_{N,\ell}\rangle=c_M\textbf{K}_{k\ell}^{(N)}+O(1)\,,
\ee
with
\be
\textbf{K}_{k\ell}^{(N)}=\frac{\ell(\ell^2-1)}{12}\Theta(k-\ell)\left(\begin{array}{c}
     N-\ell  \\
      k-\ell
\end{array}\right)\frac{(N-k)!(k+1)!}{(\ell+1)!}(2\Delta_p)_{N-k}\,,
\ee
where $\Theta$ is the discrete step function with $\Theta(0)=1$. Furthermore, it is also easy to see that $\langle \hat A_{N,k}|\hat A_{N,\ell}\rangle=O(1)$ and $\langle B_{N,k}|B_{N,\ell}\rangle=0$. As a result,
the inverse Gram matrix formed by basis $|\hat A_{N,k}\rangle$ and $|B_{N,k}\rangle$ of degree one is then given by 
\be\label{inverse}
\left(\begin{array}{cc}
  O(1)   & c_M\textbf{K}^{(N)} \\
  c_M(\textbf{K}^{(N)} )^T  & 0
\end{array}\right)^{-1}=\frac{1}{c_M}\left(\begin{array}{cc}
 0   & [(\textbf{K}^{(N)})^T]^{-1} \\
 (\textbf{K}^{(N)} )^{-1}  & 0
\end{array}\right)+O(c_M^{-2})\,,
\ee 
where the inverse matrix $(\textbf{K}^{(N)} )^{-1}$ is given by
\be
\left[(\textbf{K}^{(N)} )^{-1}\right]_{k\ell}=\frac{12(-1)^{k-\ell}(k-2)!\Theta(k-\ell)}{(N-k)!(k-\ell)!(\ell+1)!(2\Delta_p)_{N-\ell}}\,.
\ee
The inverse \eqref{inverse} determines the sector with
one non-global generator.  To show that it gives the entire contribution to the BMS block at order $c_M^{-1}$, we must also
prove that sectors with two or more non-global generators begin at $c_M^{-2}$. 
 Let $E^{\text{quot}}_{N,i}$ be the subspace at level $N$ spanned by quotient PBW states containing exactly $i$ non-global modes. We still have $\langle E^{\text{quot}}_{N,i}|E^{\text{quot}}_{N,j}\rangle=O(c_M^{\operatorname{min}(i,j)})$ as before and
\be
\langle E^{\text{quot}}_{N,i}|E^{\text{quot}}_{N,i}\rangle=c_M^i\Gamma^{\text{quot}}_{N,i}+O(c_M^{i-1})\,.
\ee
Using similar arguments of block diagonalization by induction as before, it suffices to show that $\Gamma^{\text{quot}}_{N,i}$ is nondegenerate so that states with degree greater than one contribute  at order $O(c_M^{-2})$ and can be ignored. For the degenerate case $\xi_p=0$, we can compute the matrix $\Gamma^{\text{quot}}_{N,i}$ directly and prove its nondegeneracy.  At fixed level $N$, consider descendant states of degree $i$. Their nonnegative occupation numbers $\ell_n,m_n$, with $n\geq2$, satisfy
\be
\sum_{n\geq2}(\ell_n+m_n)=i\,,\quad q\equiv\sum_{n\geq2}n(\ell_n+m_n)\leq N\,.
\ee
It is then not difficult to show that the following states
\be
|\ell,m,q\rangle=\left(\prod_{n\geq2}L^{\ell_n}_{-n}\right)\left(\prod_{n\geq2}M^{m_n}_{-n}\right)L^{N-q}_{-1}|p\rangle
\ee
form a basis of $E^{\text{quot}}_{N,i}$\,. For each occupation pattern, set
\be
\omega_{\ell,m}\equiv\prod_{n\geq2}\ell_n!m_n!\left(\frac{n(n^2-1)}{12}\right)^{\ell_n+m_n}\,.
\ee
We evaluate the coefficient of $c_M^i$ between two basis states $|\ell,m,q\rangle$ and $|\ell',m',q'\rangle$. A factor of $c_M$ can arise only from the central term in  $[L_n,M_{-n}]$ or $[M_n,L_{-n}]$ with $n\geq2$. Therefore, in order for the overlap to be of order $c_M^i$,  the two states must  have identical mode multiplicities after exchanging $L$ and $M$. Consequently, we find
\be\ba\label{inner}
\langle\ell,m,q|\ell',m',q'\rangle=&c_M^i\delta_{\ell',m}\delta_{m',\ell}\omega_{\ell,m}\langle p|L_1^{N-q}L_{-1}^{N-q}|p\rangle+O(c_M^{i-1})
\\=&c_M^i\delta_{\ell',m}\delta_{m',\ell}\omega_{\ell,m}(N-q)!(2\Delta_p)_{N-q}+O(c_M^{i-1})\,.
\ea
\ee
Here, each Kronecker symbol between occupation number sequences means equality for every $n\geq2$, i.e. $\delta_{\ell',m}=\prod_{n\geq2}\delta_{\ell'_n,m_n}$. Using \eqref{inner}, it can be shown that
\be
\det\Gamma^{\operatorname{quot}}_{N,i}\simeq\prod_{\substack{
\ell_n,m_n\ge 0\\
\sum_{n\ge 2}(\ell_n+m_n)=i\\
\sum_{n\ge 2}n(\ell_n+m_n)\le N
}}\omega_{\ell,m} a!(2\Delta_p)_a\,,\quad a=N-\sum_{n\geq2}n(\ell_n+m_n)\,.
\ee
Here, $\simeq$ denotes equality up to an overall sign which may arise from the determinant
but does not affect the nondegeneracy property.
Therefore, the determinant is nonvanishing unless
\be
\Delta_p\in\left\{-\frac{j}{2}\big|j=0,1,\cdots,N-2i-1\right\}\,.
\ee
Thus $\Gamma^{\operatorname{quot}}_{N,i}$ is nondegenerate
for every allowed $N,i$, provided
$2\Delta_p\notin\mathbb Z_{\leq0}$.
Consequently, at each fixed level, the corresponding
diagonal Gram blocks are invertible for sufficiently large
$c_M$, and their inverses begin at order $c_M^{-i}$.

Now we can proceed as in the case of $\xi_p\neq0$ to compute three point vectors. For the global sector, we have
\be\ba
\mathbf l_N^C&=\frac{\langle O_1(\infty,0)O_2(1,0)|C_N\rangle}{c_{12p}}=(\Delta_{p2,1})_N\,,\\
\mathbf r_N^C&=\frac{x^{\Delta_3+\Delta_4}e^{2\xi_3t/x}}{c_{p34}}\langle C_N|O_3(x,t)O_4(0,0)|0\rangle=x^{\Delta_p+N}(\Delta_{p3,4})_N\,.
\ea
\ee
For three point vectors containing degree one states, we define
\be
(\mathbf{l}^X_N)_k=\frac{\langle O_1(\infty,0)O_2(1,0)|X_{N,k}\rangle}{c_{12p}}\,,\quad (\mathbf{r}^X_N)_k=\frac{x^{\Delta_3+\Delta_4}e^{2\xi_3t/x}}{c_{p34}}\langle X_{N,k}|O_3(x,t)O_4(0,0)|0\rangle
\ee
with $X=A,B$. These vectors can be computed using Ward identities in a similar way as before,
and we get
\be
\ba
(\mathbf{l}^A_N)_k&=(\Delta_{p2,1})_{N-k}[\Delta_p+N-k+k\Delta_2-\Delta_1]\,,\\
(\mathbf{l}^B_N)_k&=(k-1)\xi_1(\Delta_{p2,1}+k)_{N-k}\,,\\
(\mathbf{r}^A_N)_k&=x^{\Delta_p+N}(\Delta_{p3,4})_{N-k}\left[\Delta_p+N-k+k\Delta_3-\Delta_4+k(k-1)\xi_3\frac{t}{x}\right]\,,\\
(\mathbf{r}^B_N)_k&=x^{\Delta_p+N}(k-1)\xi_3(\Delta_{p3,4}+k)_{N-k}\,.
\ea
\ee
Using the linear combination \eqref{hatA}, the three point vectors associated to the orthogonalized state $\hat A_{N,k}$ are
\be
(\mathbf{l}^{\hat A}_N)_k=(\mathbf{l}^{ A}_N)_k-\alpha_{N,k}\mathbf l^C_N\,,\quad (\mathbf{r}^{\hat A}_N)_k=(\mathbf{r}^{ A}_N)_k-\alpha_{N,k}\mathbf r^C_N\,.
\ee
The linearized BMS block is then given by
\be
\mathcal{F}^{(1)}_{\Delta_p}(x,t)=\sum_{N=2}^\infty\left[(\mathbf{l}^{\hat A}_N)^T[(\textbf{K}^{(N)})^T]^{-1}\mathbf{r}^{  B}_N+(\mathbf{l}^B_N)^T(\textbf{K}^{(N)})^{-1}\mathbf{r}^{\hat A}_N\right]\,.
\ee
Performing the summation directly, we find  $\mathcal{F}^{(1)}_{\Delta_p}$ can be expressed in terms of the global BMS block and its derivatives. More explicitly, we let $a=\Delta_{p2,1},b=\Delta_{p3,4}$ so that $\mathcal{F}^{(0)}_{\Delta_p}=x^{\Delta_p}{}_2F_1(a,b,2\Delta_p,x)$.
The linearized correction is then given by
\be
\ba\label{F1:deg}
\frac{\mathcal{F}^{(1)}_{\Delta_p}(x,t)}{12}=&\left[(\xi_3\Delta_2+\xi_1\Delta_3)S(x)+\xi_1\xi_3tS'(x)\right]\mathcal{F}^{(0)}_{\Delta_p}\\
&+(\xi_1+\xi_3)\left[\Delta_p\mathcal{F}^{(0)}_{\Delta_p}+(1-x)\log(1-x)\p_x\mathcal{F}^{(0)}_{\Delta_p}\right]\\
&+(\xi_3\Delta_{21}+\xi_1\Delta_{34})\left[1+\frac{(1-x)\log(1-x)}{x}\right]\mathcal{F}^{(0)}_{\Delta_p}\\
&+\xi_3\Delta_{34}\p_b\mathcal{F}^{(0)}_{\Delta_p}+\xi_1\Delta_{21}\p_a\mathcal{F}^{(0)}_{\Delta_p}\,,
\ea
\ee
where
\be
S(x)=-2+\left(1-\frac{2}{x}\right)\log(1-x)\,.
\ee
When the external operators are pairwise identical, only the first two lines in \eqref{F1:deg} survive.

\section{The $c^{-1}$ correction to the Virasoro block from boundary gravitons}
\label{app:virasoro-dressing}

This appendix tests the prescription of section \ref{sec:dressing} against the known
Virasoro result used in section \ref{sec:ur}. We keep the notation of \eqref{def:CFTblock}--\eqref{functions:f}:
the two identical external pairs have weights $h_1,h_3$, and the exchanged
non-vacuum primary has weight $h_p$. All weights are held fixed as
$c\to\infty$.
We first recover the leading piece
$z^{h_p}F_{h_p}(z)$ from two shadow integrals and then
contract the linear variations of their reparametrized three-point functions, keeping
the exchanged two-point function fixed as in \eqref{F1-KK}. This provides an
explicit order-$c^{-1}$ test of the construction motivated by the
non-vacuum proposal in \cite{Nguyen:2022xsw}.

Let $\widetilde O_{h_p}$ be the chiral shadow of $O_{h_p}$, with weight
$1-h_p$. For $i=1,3$ and $z_a>z_c>z_b$, define
\begin{equation}
 \begin{aligned}
 \widetilde K_i(z_a,z_b,z_c)
 &\equiv\frac{z_{ab}^{2h_i}}{C_{ii\widetilde h_p}}
 \langle O_i(z_a)\widetilde O_{h_p}(z_c)O_i(z_b)\rangle\\
 &=\left(\frac{z_{ac}z_{cb}}{z_{ab}}\right)^{h_p-1}\,.
 \end{aligned}
 \label{eq:appD-kernel}
\end{equation}
As in \eqref{OPEblock:gl}, the normalized global Virasoro OPE block $ \widetilde {\mathcal B}_{iip}^{\mathrm{gl}}$ satisfies
\begin{equation}
 \widetilde \Pi_{h_p}^{\mathrm{gl}}O_i(z_a)O_i(z_b)|0\rangle
 =C_{iih_p}z_{ab}^{-2h_i}
 \widetilde {\mathcal B}_{iip}^{\mathrm{gl}}(z_a,z_b)|0\rangle\,,
 \label{eq:appD-OPE-definition}
\end{equation}
where $\widetilde\Pi_{h_p}^{\mathrm{gl}}$ retains only the global Virasoro descendants of
$O_{h_p}$. The global Ward identities give the differential expansion,
which has the equivalent integral representation as follows \cite{Fitzpatrick:2016mtp}
\begin{equation}
 \begin{aligned}
\widetilde{ \mathcal B}_{iip}^{\mathrm{gl}}(z_a,z_b)
 &=z_{ab}^{h_p}\sum_{n=0}^{\infty}
 \frac{(h_p)_n z_{ab}^n}{n!(2h_p)_n}
 \partial_{z_b}^{n}O_{h_p}(z_b)\\
 &=\frac{\Gamma(2h_p)}{\Gamma(h_p)^2}
 \int_{z_b}^{z_a}dz_c\,\widetilde K_i(z_a,z_b,z_c)O_{h_p}(z_c)\,.
 \end{aligned}
 \label{eq:appD-OPE-integral}
\end{equation}
The leading global block, together with the appropriate normalization factor, can be written as the correlator of two global OPE blocks \footnote{The integrals are initially evaluated for $\operatorname{Re}h_p>0,0<z<1$,
and the result can be analytically continued in both $h_p$ and $z$.}
\begin{equation}
 \begin{aligned}
 z^{h_p}F_{h_p}(z)&=\langle\widetilde{\mathcal{B}}_{11p}^{\mathrm{gl}}(z_1,z_2)\widetilde{\mathcal{B}}_{33p}^{\mathrm{gl}}(z_3,z_4)\rangle\\
 &=\frac{\Gamma(2h_p)^2}{\Gamma(h_p)^4}
 \int_1^\infty dz_5\int_0^z dz_6\,
\widetilde K_1(z_1,z_2,z_5)(z_5-z_6)^{-2h_p}\widetilde K_3(z_3,z_4,z_6)\\
 &=\frac{z^{h_p}\Gamma(2h_p)^2}{\Gamma(h_p)^4}
 \int_0^1du\int_0^1dv\,
 [u(1-u)v(1-v)]^{h_p-1}(1-zuv)^{-2h_p}\,,
 \end{aligned}
 \label{eq:appD-global-integral}
\end{equation}
where 
$z_1=\infty$, $z_2=1$, $z_3=z$, $z_4=0$.
In the last line of \eqref{eq:appD-global-integral}, we have set $z_5=u^{-1}$ and $z_6=zv$. 

We now apply a conformal transformation
$O_i(z)\mapsto f'(z)^{h_i}O_i(f(z))$ to the three insertions in
\eqref{eq:appD-kernel}, while keeping the external normalization fixed. This leads to the following reparametrized kernel
\be\ba\label{eq:appD-transformed-kernel}
\widetilde K_i^f(z_a,z_b,z_c)
 ={}&\frac{z_{ab}^{2h_i}}{C_{ii\widetilde h_p}}
 f'(z_a)^{h_i}f'(z_b)^{h_i}f'(z_c)^{1-h_p}\langle O_i(f(z_a))\widetilde O_{h_p}(f(z_c))
 O_i(f(z_b))\rangle.
\ea\ee
For an infinitesimal transformation with $f(z)=z+\eta(z)$, expand
$\widetilde K_i^f=\widetilde K_i+\delta\widetilde K_i+O(\eta^2)$. Varying the three primary factors and
the coordinate differences gives
\be\ba \label{eq:appD-response}
 \frac{\delta\widetilde K_i(z_a,z_b,z_c)}{\widetilde K_i(z_a,z_b,z_c)}
 &=h_iJ_{ab}[\eta]
 +\frac{1-h_p}{2}\bigl(J_{ac}[\eta]+J_{bc}[\eta]-J_{ab}[\eta]\bigr)\,,\\
 J_{ab}[\eta]
 &=\eta'(z_a)+\eta'(z_b)
 -2\frac{\eta(z_a)-\eta(z_b)}{z_{ab}}\,.
\ea
\ee
Although $h_i$ cancels
from the undeformed kernel, it remains in the response because the
external normalization in \eqref{eq:appD-transformed-kernel} is not
transformed.
Following \eqref{F1-KK}, the quantity we are about to compute is
\be\ba\label{eq:appD-correction-prescription}
\delta G_{h_p}(c;z)
\equiv{}&
\frac{\Gamma(2h_p)^2}{z^{h_p}\Gamma(h_p)^4}
\int_1^\infty dz_5\int_0^z dz_6\,
(z_5-z_6)^{-2h_p}
\langle\delta\widetilde K_1(\infty,1,z_5)
\delta\widetilde K_3(z,0,z_6)\rangle
\\
={}&
\frac{\Gamma(2h_p)^2}{\Gamma(h_p)^4}
\int_0^1du\int_0^1dv\,
[u(1-u)v(1-v)]^{h_p-1}(1-zuv)^{-2h_p}\\
&\quad\times
\frac{\langle\delta\widetilde K_1(\infty,1,u^{-1})
\delta\widetilde K_3(z,0,zv)\rangle}
{\widetilde K_1(\infty,1,u^{-1})
\widetilde K_3(z,0,zv)}\,,
\ea
\ee
where in the second equality we have used $u=z_5^{-1}$ and $v=z_6/z$, as in \eqref{eq:appD-global-integral}.
The correlator of reparametrization modes on the plane is given by  \cite{Haehl:2019eae,Anous:2020vtw,Nguyen:2021jja}
\be
\langle\eta(z_a)\eta(z_b)\rangle=\frac{6}{c}z_{ab}^2\log z_{ab}\,.
\ee
Consequently, the correlator of two $J$ operators is computed to be
\begin{equation}
 \langle J_{ab}[\eta]J_{cd}[\eta]\rangle
 =\frac{12}{c}\left[-2+
 \left(1-\frac{2}{\chi}\right)\log(1-\chi)\right]\,,
 \qquad \chi=\frac{z_{ab}z_{cd}}{z_{ac}z_{bd}}\,.
 \label{eq:appD-JJ}
\end{equation}
Substituting \eqref{eq:appD-JJ} into the two linear variations in
\eqref{eq:appD-response} gives
\begin{equation}
\begin{aligned}
&\frac{\langle\delta\widetilde K_1(\infty,1,u^{-1})
\delta\widetilde K_3(z,0,zv)\rangle}
{\widetilde K_1(\infty,1,u^{-1})\widetilde K_3(z,0,zv)}
\\
&=\frac{12}{c}\Bigg\{
h_1h_3\left[-2+\left(1-\frac{2}{z}\right)\log(1-z)\right]
\\
&\quad+h_1(h_p-1)\left[
1+\frac{v(1-z)}{z(1-v)}\log(1-z)
+\frac{1-2v+zv^2}{zv(1-v)}\log(1-zv)\right]
\\
&\quad+h_3(h_p-1)\left[
1+\frac{u(1-z)}{z(1-u)}\log(1-z)
+\frac{1-2u+zu^2}{zu(1-u)}\log(1-zu)\right]
\\
&\quad+(h_p-1)^2\Bigg[
-\frac12-\frac{1}{2zuv(1-u)(1-v)}
\Bigl\{u^2v^2(1-z)^2\log(1-z)
\\
&\qquad\qquad
+v^2(1-zu)(1-2u+zu)\log(1-zu)
\\
&\qquad\qquad
+u^2(1-zv)(1-2v+zv)\log(1-zv)
\\
&\qquad\qquad
+\bigl[1-2u-2v+4uv-z(2-z)u^2v^2\bigr]
\log(1-zuv)\Bigr\}\Bigg]\Bigg\}\,,
\end{aligned}
\label{eq:appD-contraction-position}
\end{equation}
 Expanding the logarithms as
$\log(1-w)=-\sum_{n\geq1}\frac{w^n}{n}$ and collecting powers of $z$, we can rewrite \eqref{eq:appD-contraction-position} in the following form
\begin{equation}
\begin{aligned}
&\frac{\langle\delta\widetilde K_1(\infty,1,u^{-1})
\delta\widetilde K_3(z,0,zv)\rangle}
{\widetilde K_1(\infty,1,u^{-1})\widetilde K_3(z,0,zv)}
\\
&=\frac{12}{c}\sum_{s=2}^{\infty}
\frac{z^s}{s(s^2-1)}
\left[h_1(s-1)+(h_p-1)
\frac{u\bigl[1-su^{s-1}+(s-1)u^s\bigr]}{1-u}\right]
\\
&\qquad\qquad\qquad\times
\left[h_3(s-1)+(h_p-1)
\frac{v\bigl[1-sv^{s-1}+(s-1)v^s\bigr]}{1-v}\right]\,.
\end{aligned}
\label{eq:appD-contraction-series}
\end{equation}
The constant and linear powers of $z$ cancel in the logarithmic
expression, so the sum starts at $s=2$.

We now insert \eqref{eq:appD-contraction-series} into
\eqref{eq:appD-correction-prescription} and evaluate the integrals. Expanding the remaining
factor as
\begin{equation}
 (1-zuv)^{-2h_p}
 =\sum_{n=0}^{\infty}\frac{(2h_p)_n}{n!}z^nu^nv^n
\label{eq:appD-pairing-series}
\end{equation}
separates the $u$ and $v$ integrals. Performing these integrals and
collecting the total power $q=n+s$ of $z$ yields
\begin{equation}
 \delta G_{h_p}(c;z)
 =\frac{12}{c}\sum_{q=2}^{\infty}z^q
 \sum_{s=2}^{q}
 \frac{B_{q,s}(h_1)B_{q,s}(h_3)}
 {s(s^2-1)(q-s)!(2h_p)_{q-s}}\,,
 \label{eq:appD-integrated-series}
\end{equation}
where each coefficient is the evaluated integral
\begin{equation}
\begin{aligned}
 B_{q,s}(h_i)
 &\equiv\frac{\Gamma(2h_p)(2h_p)_{q-s}}{\Gamma(h_p)^2}
 \int_0^1du\,u^{h_p+q-s-1}(1-u)^{h_p-1}
 \\
 &\qquad\times\left[h_i(s-1)+(h_p-1)
 \frac{u\bigl[1-su^{s-1}+(s-1)u^s\bigr]}{1-u}\right]
 \\
 &=[h_i(s-1)+h_p+q-s](h_p)_{q-s}
-[h_p(s+1)+q-s]
 \frac{(2h_p)_{q-s}}{(2h_p)_q}(h_p)_q\,.
\end{aligned}
\label{eq:appD-integrated-coefficient}
\end{equation}
The coefficients and double series in
\eqref{eq:appD-integrated-coefficient} and
\eqref{eq:appD-integrated-series} coincide with those in
\cite{Bombini:2018jrg}. Using their resummation, we eventually obtain
\begin{equation}
 \delta G_{h_p}(c;z)
 =\frac{12}{c}
 \left[h_1h_3f_a(z)+(h_1+h_3)f_b(z)+f_c(z)\right]\,,
 \label{eq:appD-correction-result}
\end{equation}
where $f_a,f_b,f_c$ are the functions in \eqref{functions:f}.
Thus the contraction of the two linearized kernels reproduces the
order-$c^{-1}$ term in
$G_{h_p}(c;z)=F_{h_p}(z)+\delta G_{h_p}(c;z)+O(c^{-2})$,
with the normalization used in section~\ref{sec:ur}.

\bibliographystyle{JHEP}
\bibliography{bms_block}

\end{document}